\documentclass[11pt]{article}

\usepackage{mathtools}
\usepackage{amsmath,amsfonts,amssymb}
\usepackage{latexsym}
\usepackage{dsfont}
\usepackage{shuffle}
\usepackage{slashed}
\usepackage{cancel}
\usepackage{bbold}
\usepackage{mathbbol}

\usepackage{epsfig}
\usepackage{graphicx}
\usepackage{epstopdf}
\usepackage{tikz}
\usetikzlibrary{decorations.pathmorphing}
\tikzset{snake it/.style={decorate, decoration=snake}}

\usepackage{xcolor}
\usepackage{color}
\usepackage{soul}

\usepackage{scalerel}
\usepackage{stackrel}
\usepackage{authblk}
\usepackage[hang,flushmargin]{footmisc}
\usepackage[hidelinks]{hyperref}
\definecolor{blue}{RGB}{25, 0, 149}
\hypersetup{
 colorlinks,
 linkcolor={blue},
 citecolor={blue},
 urlcolor={blue}
}

\usepackage{hyperref}
\usepackage{cite}

\usepackage{comment} 
\allowdisplaybreaks

\def\hybrid{
        \topmargin -20pt
        \oddsidemargin 0pt
        \headheight 0pt \headsep 0pt
        \textwidth 6.25in 
        \textheight 9.5in 
        \marginparwidth .875in
        \parskip 5pt plus 1pt \jot = 1.5ex}
\hybrid
\csname @addtoreset\endcsname{equation}{section}

\newcommand{\be}{\begin{equation}}
\newcommand{\ee}{\end{equation}}
\newcommand{\ba}{\begin{equation} \begin{aligned}}
\newcommand{\ea}{\end{aligned} \end{equation}}
\def\la{\langle}
\def\ra{\rangle}
\def\cC{{\cal C}}
\def\cB{{\cal B}}

\def\cD{\mathcal{D}}
\def\cF{{\cal F}}
\def\cO{{\cal O}}
\def\cA{{\cal A}}

\def\cN{{\cal N}}

\def\cP{{\cal P}}

\def\cV{{\cal V}}

\def\del{\partial}
\def\l{\langle}
\def\r{\rangle}

\def\B{\square}

\allowdisplaybreaks

\def\udot{\dot{\phantom{a}}\!}
\def\bpm{\begin{pmatrix}}
\def\epm{\end{pmatrix}}
\newcommand{\bra}[1]{\langle#1\rvert}
\newcommand{\ket}[1]{\lvert#1\rangle}

\def \diagramThree{
\begin{figure}[ht!]
\centering
\begin{tikzpicture}[scale=1.75]
    \node at (-0.5,0) {$\cA_{123}=$};
    \node at (0,0) {x};
    \node at (0,0.3) {$1$};
    \draw (0,0)--(4,0);
    \draw (2.5,1.25)--(2.5,0);
    \node at (2.3,1.25) {$2$};
    \node at (2.5,-0.1) {\small{$0$}};
    \node at (4,0) {x};
    \node at (4,0.3) {$3$};
\end{tikzpicture}
\caption{Diagrammatic representation of the color-ordered three-gluon amplitude $\cA_{123}$. The straight line denotes the insertion of an unintegrated vertex operator at $\tau=0$. The asymptotic endpoints of the worldline are depicted by a cross.}
\label{fig1}
\end{figure}
}

\def \diagramFourSmart{
\begin{figure}[ht!]
\centering
\begin{tikzpicture}[scale=1.75]
    \node at (-0.5,0) {$\cA_{1234}=$};
    \node at (0,0) {x};
    \node at (0,0.3) {$2$};
    \draw (0,0)--(4,0);
    \draw[snake it] (1.5,-1.25)--(1.5,0);
    \node at (1.3,-1.25) {$1$};
    \draw (2.5,1.25)--(2.5,0);
    \node at (2.3,1.25) {$3$};
    \node at (1.5,0.1) {\small{$\tau$}};
    \node at (2.5,-0.1) {\small{$0$}};
    \node at (4,0) {x};
    \node at (4,0.3) {$4$};
\end{tikzpicture}
\caption{Diagrammatic representation of the color-ordered four-gluon amplitude $\cA_{1234}$ with the smart choice for the worldline. The wavy line denotes the insertion of an integrated vertex operator, while, as before, the straight line denotes the one fixed by translation invariance.}
\label{fig2}
\end{figure}
}

\def \diagramFourDumb{
\begin{figure}[ht!]
\centering
\begin{tikzpicture}[scale=1]
    \node at (-1,0) {$\cA_{1234}=$};
    \node at (0,0) {x};
    \node at (0,0.3) {$1$};
    \draw (0,0)--(4,0);
    \draw[snake it] (1.5,1.25)--(1.5,0);
    \node at (1.3,1.25) {$2$};
    \draw (2.5,1.25)--(2.5,0);
    \node at (2.3,1.25) {$3$};
    \node at (1.5,-0.2) {\small{$\tau$}};
    \node at (2.5,-0.2) {\small{$0$}};
    \node at (4,0) {x};
    \node at (4,0.3) {$4$};

    \node at (4.5,0) {$+$};

    \node at (5,0) {x};
    \node at (5,0.3) {$1$};
    \draw (5,0)--(9,0);
    \draw (6.5,1.25)--(7,0)--(7.5,1.25);
    \node at (6.3,1.25) {$2$};
    \node at (7.7,1.25) {$3$};
    \node at (7,-0.2) {\small{$0$}};
    \node at (9,0) {x};
    \node at (9,0.3) {$4$};

    \node at (9.5,0) {$+$};

    \node at (10,0) {x};
    \node at (10,0.3) {$1$};
    \draw (10,0)--(14,0);
    \draw (12,0.7)--(12,0);
    \draw (11.5,1.25)--(12,0.7)--(12.5,1.25);
    \node at (11.3,1.25) {$2$};
    \node at (12.7,1.25) {$3$};
    \node at (12,-0.2) {\small{$0$}};
    \node at (14,0) {x};
    \node at (14,0.3) {$4$};
\end{tikzpicture}
\caption{Diagrammatic representation of the color-ordered four-gluon amplitude $\cA_{1234}$ with the not-so-smart choice for the worldline. As before, the wavy line denotes the insertion of an integrated vertex operator, while the straight lines denote the insertion of unintegrated vertex operators at $\tau=0$.}
\label{fig3}
\end{figure}
}

\begin{document}

\title{
\begin{flushright}
\vspace{-0.5cm}
{\small HU-EP-25/28}
\end{flushright}
\vspace{1cm}
\bf Gluon amplitudes in first quantization}

\author[a,d]{Fiorenzo Bastianelli\,\footnote{E-mail: fiorenzo.bastianelli@unibo.it}}
\author[b]{Roberto Bonezzi\,\footnote{E-mail: roberto.bonezzi@physik.hu-berlin.de}}
\author[c,d]{Olindo Corradini\,\footnote{E-mail: olindo.corradini@unimore.it}}
\author[a,d]{Filippo Fecit\,\footnote{E-mail: filippo.fecit2@unibo.it}}

\affil[a]{\small\it Dipartimento di Fisica e Astronomia ``Augusto Righi", \protect\\ Universit\`a di Bologna, Via Irnerio 46, I-40126, Bologna, Italy\protect\\ \phantom{ciccio}}

\affil[b]{{\small\it  Institute for Physics,\protect\\ Humboldt University Berlin,
 Zum Gro\ss en Windkanal 2, D-12489 Berlin, Germany}\protect\\ \phantom{ciccio}}

\affil[c]{{\small\it Dipartimento di Scienze Fisiche, Informatiche e Matematiche, \protect\\
Universit\`a degli Studi di Modena e Reggio Emilia, 
Via Campi 213/A, I-41125 Modena, Italy}\protect\\ \phantom{ciccio}}

\affil[d]{{\small\it INFN, Sezione di Bologna,  Via Irnerio 46, I-40126 Bologna, Italy}}

\date{}

\maketitle
 
\bigskip\bigskip
\begin{center} 
\textbf{Abstract}

\end{center} 
\begin{quote}
We compute tree-level gluon amplitudes as worldline correlators of vertex operators in a bosonic spinning particle model. In this framework, the particle’s position degrees of freedom are extended by complex bosonic variables that encode its spin. In the free theory, the model exhibits a first-class constraint algebra whose gauging ensures the unitarity of the quantum theory. This algebra is a contraction of the  $sl(2,\mathbb{R})$ algebra, which is by itself a subalgebra of the Virasoro algebra. In string theory, gauging the Virasoro algebra plays a similar role.
Our model admits a consistent truncation to describe a pure spin-1 particle. Non-abelian interactions are introduced by using BRST techniques, which allow us to extract the vertex operators of the theory as suitable deformations of the BRST charge. BRST invariance is central to ensuring the consistency of the tree-level amplitudes analyzed in this work. We discuss connections with similar worldline constructions and comment on the potential relevance of this framework for uncovering the structures underlying the double-copy program in gauge and gravitational theories.
\end{quote}

\vfill
\setcounter{footnote}{0}

\newpage

\tableofcontents

\vspace{5mm}

\section{Introduction}
We present a first-quantized approach to computing tree-level gluon amplitudes using BRST methods and worldline path integrals, building on the recent treatment of Ref. \cite{Bonezzi:2024emt} and the seminal work of Ref. \cite{Dai:2008bh}.
Our framework is based on a bosonic particle that, along with the usual geometric degrees of freedom describing the particle’s spacetime position, introduces bosonic partners that encode the spin degrees of freedom. The particle has a symmetry algebra that relates the spacetime coordinates to some suitable bosonic partners, which are the bosonic analogs of supersymmetric coordinates. Such algebra is sometimes referred to as ``bosonic supersymmetry".

Traditionally, relativistic spinning particles are described using Grassmann-valued degrees of freedom, related to the coordinates via supersymmetry (see Refs. \cite{Berezin:1976eg, Barducci:1976qu, Brink:1976sz}, with extensions to higher spin discussed in Refs. \cite{Gershun:1979fb, Howe:1988ft}). The alternative use of bosonic variables to treat particles with integer spin has a long history, mainly rooted in string theory \cite{Bengtsson:1986ys, Henneaux:1987cp, Bouatta:2004kk, Hallowell:2007qk}. However, it has been explored only sporadically in the context of worldline path integrals, as in \cite{Bastianelli:2009eh}. It has seen renewed interest in recent works, specifically in the context of Yang-Mills particles \cite{Bonezzi:2024emt}, which is the focus of our present endeavor, 
in the new discussion of a charged massive spin-1 
particle \cite{Bastianelli:2025khx},
and in the study of gravitational radiation from spinning bodies \cite{Haddad:2024ebn, Mogull:2025cfn}.

The traditional formulation of the spin-1 particle involves complex fermionic partners $(\psi^\mu, \bar\psi^\mu)$ associated with the particle coordinates $x^\mu$, along with a gauged $N=2$ worldline supersymmetry. This structure leads to a first-class constraint algebra, whose role is essential for establishing unitarity and ensuring that only the massless spin-1 sector is propagated. This set-up was thoroughly analyzed in Ref. \cite{Howe:1989vn}, where it was also shown that introducing a quantized Chern-Simons coupling on the worldline enables the propagation of a massless spin-1 particle -- and more generally the quanta of antisymmetric gauge fields -- in arbitrary space-time dimensions.
However, it was also pointed out that challenges arise when coupling the model to background fields. Specifically, coupling to electromagnetism breaks the local $N=2$ supersymmetry, rendering the theory inconsistent. In contrast, coupling to background gravity is viable and effectively employed in Refs. \cite{Bastianelli:2005vk, Bastianelli:2005uy} to study the gravitational one-loop effective action induced by vector and antisymmetric tensor fields.

A breakthrough came with the work of Ref. \cite{Dai:2008bh}, which demonstrated that coupling to a Yang-Mills background is feasible. Using BRST techniques, it was shown that a nilpotent BRST charge can be maintained in the sector of the Hilbert space corresponding to the non-abelian spin-1 degrees of freedom, provided the Yang-Mills background satisfies its equations of motion. Tree-level amplitudes and the one-loop beta function were thus computed within this worldline approach.

Here, we aim to pursue a similar analysis using the particle model based on bosonic oscillators. In Ref. \cite{Bonezzi:2024emt}, one of us demonstrated that the model describes excitations of arbitrary integer spin, and that a restriction to the spin-1 sector admits a coupling to the Yang-Mills background, yielding an alternative worldline representation of a non-abelian spin-1 particle. The analysis was carried out in the operatorial BRST framework. In this work, we develop the corresponding path integral formulation, compute color-ordered tree-level amplitudes up to four points, and address several technical aspects that are useful to explore the rich structure that emerges in the first-quantized picture.

In particular, we consider the path integral along the infinite line, which we use to compute color-ordered tree-level amplitudes. Let us point out that the infinite line geometry implements automatically the LSZ reduction \cite{Laenen:2008gt,Bonocore:2020xuj,Mogull:2020sak,Bonezzi:2025iza}, and allows one to directly obtain on-shell amplitudes from the worldline path integral. The infinite line constitutes the main worldline that connects two external points and can be chosen arbitrarily. It describes asymptotically free vector particles.
Vertices linking the worldline to the remaining external (asymptotic on-shell) particle states 
emerge in part from the perturbative expansion of the particle action, leading to insertions of vertex operators. Insertion of
additional composite vertex operators, whose structure shall be discussed in detail, are needed to obtain the complete BRST-invariant color-ordered amplitudes.

The BRST symmetry is used to investigate, in general terms, the structure and properties of the vertex operators -- both integrated and unintegrated -- that contribute to the construction of the color-ordered amplitudes. It serves as the guiding principle for identifying whether additional vertex operators, which we refer to as ``pinch operators", must be inserted to restore BRST invariance when it is broken in perturbatively constructed tree-level amplitudes. The role of such composite pinch operators is to insert entire subtrees into the main worldline. These generalized vertex operators can be obtained either from the Bern-Kosower pinching rules \cite{Bern:1990cu,Bern:1990ux,Bern:1991aq,Bern:1991an,Schubert:2001he} or, in a second quantized framework, by Berends-Giele recursion relations \cite{Berends:1987me}. In this work, we argue that the subtrees (or multiparticle fields) can be obtained directly from a path integral on a semi-infinite line, thereby generalizing a similar construction proposed in \cite{Bonezzi:2025iza} for scalar amplitudes.

Although the overall framework outlined above employs a bosonic particle model, the structure is conceptually similar to the one proposed in \cite{Dai:2008bh}, where a particle model with fermionic oscillators
was used. Here, we further explore the role played by the BRST symmetry. Namely, we show how the decoupling of BRST-exact operators in correlation functions corresponds to the target-space Ward identities, similarly to what occurs in string theory. This analysis also justifies the need for multiparticle pinch operators as a necessary condition to restore BRST invariance. 
Technically, we find it useful to evaluate the perturbative expansion of the path integral, 
and the execution of checks of the BRST Ward identities,
in terms of worldline operator product expansions (OPEs), as in \cite{Dai:2008bh} and more recently in \cite{Du:2023nzo, Du:2024rkf, Ajith:2024fna}. They help organize the key algebraic properties emerging from the path integral, which are closely related to the homotopy algebra framework employed in \cite{Nutzi:2018vkl,Arvanitakis:2019ald,Macrelli:2019afx,Lopez-Arcos:2019hvg,Bonezzi:2023xhn,Bonezzi:2024emt,Bonezzi:2024fhd}.

The rest of the paper is organized as follows. In Section \ref{sec2}, we introduce the free bosonic spinning particle model, set up the path integral on the finite line, and review its consistent coupling to a Yang-Mills background using BRST techniques. We show how the vertex operators naturally emerge from the BRST analysis. In Section \ref{sec3}, we compute tree-level color-ordered amplitudes, obtained as correlation functions of vertex operators inserted along open worldlines. We begin with the simpler three-gluon amplitude, then proceed to the four-gluon amplitude, comparing different worldline choices. This demonstrates the independence of the result from the choice of a specific worldline, and also highlights that selecting a worldline connecting two of the least color-adjacent gluons leads to a simpler computation. Section \ref{sec4} is dedicated to the analysis of the Ward identities satisfied by the amplitudes computed. Section \ref{sec5} contains our conclusions and outlook. Technical details regarding the phase-space path integral are presented in Appendix \ref{Appendix:A}.


\section{The bosonic spinning particle in a Yang-Mills background} \label{sec2}

\subsection{The free spinning particle} \label{sec2.1}
We begin by outlining the structure of the free worldline action of the bosonic spinning particle that we wish to use.
A natural starting point is the symplectic term
\begin{equation}
S_{\rm symp}=\int_0^1 d\tau\,\Big[p_\mu\dot x^\mu-i\,\bar\alpha_\mu\dot\alpha^\mu\Big]\;. 
\end{equation}
It defines the canonical Poisson brackets for the particle's phase-space coordinates and momenta $(x^\mu,p_\mu)$ joined by the bosonic oscillator pair $(\alpha^\mu,\bar\alpha_\mu)$. 
Here, $\mu$ is a Lorentz index in the 
$D$-dimensional Minkowski target space, and  $\bar\alpha^\mu$ is the complex conjugate of $\alpha^\mu$. 
The elementary Poisson brackets are given by
 \be
 \{x^\mu, p_\nu\} = \delta^\mu_\nu 
 \;, \qquad 
 \{\alpha^\mu, \bar \alpha_\nu\} = i \delta^\mu_\nu \;.
  \ee
  One may notice that the bosonic oscillators resemble a worldline analogue of the open string modes 
$\alpha^\mu_{\pm1}$.
We now define the following conserved phase-space functions
\begin{equation}
H = \frac{1}{2}\,p^2 \;, \quad
L = \alpha^\mu p_\mu \;, \quad
\bar L = \bar\alpha^\mu p_\mu\;,
\end{equation}
which form a closed algebra under Poisson brackets
\begin{equation} \label{constraint algebra}
\{L,\bar L\}= 2i H\;,
\end{equation}
while other independent brackets vanish. This algebra is first-class and bears a close resemblance to a supersymmetry algebra, except that $L$ and $\bar L$ are bosonic rather than fermionic. 
It corresponds to a contraction of the $sl(2, \mathbb{R})$ algebra \cite{Bengtsson:1986ys,Bouatta:2004kk}.
The charges ($H, L, \bar L)$ generate worldline symmetries: $H$ is the generator of $\tau$-translations, while 
$L$ and $\bar L$  mix the spacetime coordinates 
$x^\mu$ with the oscillator variables $\alpha^\mu$ and $\bar \alpha^\mu$, respectively.  
Their structure is reminiscent of the Virasoro generators $L_0$ and $L_{\pm1}$ 
 of the bosonic open string.

Upon quantization, the states of the theory are interpreted as those corresponding to massless particles, with the spin degrees of freedom carried by the oscillator variables $\alpha^\mu$. 
 To this end, one must impose the mass-shell constraint $p^2=0$ by gauging the Hamiltonian $H$, 
 and eliminate remaining unphysical states by gauging the ``Virasoro-like" generators  $L$ and $\bar L$.
This gauging plays a role analogous to the gauging of the Virasoro algebra in string theory, where lightcone oscillator modes must be removed to avoid negative-norm states.

The full worldline action, incorporating these gauge symmetries, is given by
\begin{equation}
S=\int_0^1 d\tau\,\Big[p_\mu\dot x^\mu-i\,\bar\alpha^\mu\dot\alpha_\mu-e\,H-\bar u\,L-u\,\bar L\Big] \;,
\label{act-1}
\end{equation}
where $e(\tau)$ is the einbein enforcing the mass-shell condition $H=0$,
and  $u(\tau)$ and $\bar u(\tau)$
are complex Lagrange multipliers enforcing $\bar L= L=0$.
The action is invariant under local worldline reparametrizations and the bosonic analog 
of supersymmetry, with infinitesimal transformations
\ba
&\delta x^\mu=\epsilon\,p^\mu+\xi\,\bar\alpha^\mu+\bar\xi\,\alpha^\mu\;,  
\hskip .8cm 
\delta p_\mu=0\;,
\\
&\delta\alpha^\mu=i\,\xi\,p^\mu\;,
\hskip 2.9cm 
\delta\bar\alpha^\mu=-i\,\bar\xi\,p^\mu\;,
\\
&\delta e=\dot\epsilon+2i\,u\,\bar\xi-2i\,\bar u\,\xi\;, 
\hskip 1.2cm 
\delta u=\dot\xi\;, 
\hskip 2cm 
\delta\bar u=\dot{\bar\xi}\;,
\label{2.6}
\ea
where $\epsilon(\tau)$ is the local parameter for worldline reparametrizations, while $\xi(\tau)$
and $\bar\xi(\tau)$ are complex gauge parameters for the bosonic symmetry.

The quantum theory corresponding to this action has been analyzed 
in \cite{Bonezzi:2024emt} to identify covariantly the physical sector 
of the Hilbert space, using both the Dirac quantization method 
and the more general BRST approach. 
The resulting spectrum was found to include an infinite tower of bosonic spinning particles. 
This fact emerges by recalling that the quantized phase-space coordinates obey the commutation relations
\be
 [x^\mu, p_\nu] = i \delta^\mu_\nu 
 \;, \qquad 
 [\bar \alpha_\nu, \alpha^\mu ] =  \delta_\nu^\mu\;,
 \label{comm-rel}
  \ee
with the $\alpha^\mu$ oscillators that act as creation operators on the Fock vacuum $|0\ra$.
The covariant analysis carried out in \cite{Bonezzi:2024emt} shows that, 
at occupation number $s$, defined 
according to the number operator $N= \alpha^\mu  \bar \alpha_\mu$, the Hilbert space contains particle's excitations of spin 
  $s, s-2,\cdots, 0$ for even $s$, and $s, s-2,\cdots, 1$ for odd $s$. 
  A quick way to see this is by adopting the light-cone gauge, in which the bosonic creation operators $\alpha^i$
 carry only transverse indices $i=1,\cdots, D-2$. At level  $s$,  the action of $s$ such creation operators on the Fock vacuum
 yields states corresponding to a totally symmetric wavefunction $\phi_{{i_1}\cdots{i_s}}(x)$, which includes all possible traces. 
 Factorizing iteratively the traces shows that  the wavefunction naturally accommodates spin states $s,s-2,\cdots$ 
down to either 0 or 1, depending on whether  $s$ is even or odd.
 
In the rest of this section, we construct a path integral for this model.
One can anticipate some difficulties in the construction. 
Considering a finite line, the gauge transformation laws in \eqref{2.6} suggest that all the gauge fields
$(e, u, \bar u)$ should each acquire a modulus,
with the set of moduli collectively denoted by  $ (2T, u_0, \bar u_0)$, where $T$ is the usual Schwinger proper time. 
After gauge-fixing the gauge fields to 
the corresponding moduli, i.e. setting
$(e, u, \bar u) =(2T, u_0, \bar u_0)$, one finds that 
the action \eqref{act-1} reduces to
\begin{equation} 
S_{\mathrm{gf}}=\int_0^1 d\tau\,\Big[p\cdot\dot x-i\,\bar\alpha\cdot\dot\alpha-Tp^2-\bar u_0\, p\cdot\alpha -u_0\,p\cdot\bar \alpha\Big] \;.
\label{act-2}
\end{equation}
An integration over the moduli, with  $T\in \mathbb{R}^+$ 
and $u_0, \bar u_0 \in \mathbb{R}$,
produces for constant $p, \alpha, \bar \alpha$ 
\begin{equation}
\frac{1}{p^2}\ \delta(p\cdot\alpha)\ \delta(p\cdot\bar \alpha)\;,
\end{equation}
which, on top of delivering the massless propagator, seems to enforce the transversality of the oscillators. However, we are going to argue that, for our purposes, we do not need to consider the moduli for $u$ and $\bar u$. 
The reason is that we intend to restrict the model to obtain the propagation of particles of fixed spin $s$ only, and in particular $s=1$, to describe massless gluons.
To achieve such a projection, it is useful to gauge the global $U(1)$ symmetry, generated by the number charge $N=\alpha^\mu \bar\alpha_\mu$,
and add a Chern-Simons coupling,  to obtain a constraint that projects physical states to have
occupation number $s$ in the quantum setting.

Therefore, we gauge the global $U(1)$ symmetry 
\ba
\alpha' &= e^{i \phi} \alpha \;,    
\qquad
 \bar \alpha'= e^{-i\phi} \bar \alpha\;,  \\
u' &= e^{i\phi}u \;,
\qquad  
  \bar u'= e^{-i\phi} \bar u\;, \\
\ea
and add a Chern-Simons coupling with charge $q$, left unspecified for the moment, to be related to the spin $s$. We arrive at the modified action
\begin{equation}
S=\int_0^1 d\tau\,\Big[p_\mu\dot x^\mu-i\,\bar\alpha_\mu\dot\alpha^\mu-e\,\frac 12 p^2 -\bar u \, \alpha^\mu p_\mu 
-u\ \bar\alpha^\mu p_\mu- a \, 
( \alpha^\mu \bar\alpha_\mu - q)
\Big]\;.
\label{act-3}
\end{equation}
Here $(e,\bar u, u, a)$ are the full set of worldline gauge fields that gauge the charges $(H, L, \bar L, J)$ defined by
\begin{equation}
H=\frac12\,p^2\;,
\quad 
L=\alpha^\mu p_\mu \;,
\quad 
\bar L=\bar\alpha^\mu p_\mu\;, 
\quad 
 J=\alpha^\mu \bar\alpha_\mu -q \;. 
 \label{charges}
\end{equation} 
They satisfy the following Poisson bracket algebra 
\begin{equation}
\{L,\bar L\}= 2i H \;, \quad   \{J, L\}= -i L \;, \quad  \{J,\bar L\}= i \bar L\;,
\end{equation}
while other brackets vanish. It is a first-class algebra and thus its gauging is consistent.
Denoting by  $(\epsilon, \bar \xi, \xi, \phi)$ the gauge parameters 
for the gauge fields above, we get the following infinitesimal transformation rules for the gauge symmetries of the action \eqref{act-3}

\ba
&\delta x^\mu=\epsilon\,p^\mu+\xi\,\bar\alpha^\mu+\bar\xi\,\alpha^\mu\;,  
\hskip .8cm 
\delta p_\mu=0\;,
\\
&\delta\alpha^\mu=i\,\xi\,p^\mu+i \,\phi\, \alpha^\mu    
\;,
\hskip 1.61cm 
\delta\bar\alpha^\mu=-i\,\bar\xi\,p^\mu -i \,\phi \,\bar \alpha^\mu\;,
\\
&\delta e=\dot\epsilon+2i\,u\,\bar\xi-2i\,\bar u\,\xi\;, 
\hskip 1.2cm 
\delta a=\dot{\phi}\;,
\\
&\delta u=\dot\xi -i\, a\, \xi +i\, \phi \,u
\;,  
\hskip 1.7cm 
\delta\bar u=\dot{\bar\xi} +i\, a\, \bar \xi -i\, \phi \, \bar u \;.
\ea

Notice that the $U(1)$ gauge field $a$ admits a modulus, $a(\tau)=\theta$,
which in turn prevents the emergence of moduli for $u$ and $\bar u$. 
A quantum massless spin-$s$ particle is obtained by quantizing precisely this action. Before proceeding with the path integral quantization, let us clarify the relation between the Chern-Simons coupling $q$ and the spin $s$. This relation generally depends on the quantization scheme that is adopted, which we make explicit in the course of our discussion.

Using canonical quantization, we recall that the quantum operators satisfy the commutation relations in \eqref{comm-rel} with $\alpha$ acting as creation operator. Then, 
of the charges in \eqref{charges}, 
only $J$ exhibits potential ordering ambiguities. We resolve them by adopting a symmetric ordering prescription for the oscillators. This is our chosen quantization scheme and yields the quantum operator
\begin{align}
J &= \frac{1}{2} (\alpha^\mu \bar{\alpha}_\mu + \bar{\alpha}_\mu \alpha^\mu) - q
\cr
&= \alpha^\mu \bar{\alpha}_\mu + \frac{D}{2} - q
\cr
&= N - s\;,
\label{quantumJ}
\end{align}
where we have used the commutation relations of the oscillators, recognized the number operator
\be
N = \alpha^\mu \bar{\alpha}_\mu 
\label{2.16}
\ee
which counts the occupation number in the Fock space of states, and related the Chern-Simons coupling $q$ to the real parameter $s$ by
\be
q = \frac{D}{2} + s \;.
\ee
Imposing the constraint {\it à la} Dirac to select a generic physical state $|\phi\rangle$,
\be
J |\phi\rangle = 0 \qquad \Rightarrow \qquad N |\phi\rangle = s |\phi\rangle\;,
\ee
shows that physical states must have occupation number $s$. This corresponds to having maximal spin $s$.
Later, we specialize to $s = 1$, as we intend to focus on spin-1 particles. However, for the time being, we keep $s$ arbitrary and observe that it must be a non-negative integer in order to find nontrivial solutions to the constraint equation, and allow for a nontrivial quantum theory. This illustrates the well-known fact that Chern-Simons couplings are quantized.

Now, let us proceed with the worldline path integration and derive the propagator for a fixed spin-$s$ particle (together with its lower spin tail, as discussed above).
We consider the path integral on the finite line, with fixed initial and final states chosen to be
 momentum eigenstates for the $(x,p)$-variables and coherent states for the bosonic oscillators. This leads to 
the path integral representation for the particle propagator in momentum space
 \begin{equation}
_{out}\la p', \alpha | p, \bar \alpha \ra_{in}= \int \frac{DX D G}{\text{Vol(Gauge)}}\, 
e^{iS}\;,
 \end{equation}
which uses the action in \eqref{act-3}. 
On the left-hand side, we have indicated by $p$ and $p'$ the momentum eigenstates and $\alpha $ and $\bar\alpha$ the coherent states for the oscillators
(no confusion should arise with this notation).
Also, we have collectively denoted the dynamical fields by
 by $X=(x^\mu, p_\mu, \alpha^\mu, \bar \alpha_\mu)$
and the gauge fields by $G=(e, \bar u,u, a)$. 
Note that, to impose boundary conditions on $p$, 
one must add boundary terms to the action. Their final effect is to change the symplectic term $p_\mu\dot x^\mu$
  into $- \dot p_\mu x^\mu$, see Appendix \ref{Appendix:A}.

At this stage, we gauge-fix the gauge fields to their respective moduli, setting  $(e, \bar u, u, a) = (2T, 0,0, \theta)$, where $T$ and $\theta$ denote the moduli.
The latter must be integrated over a suitable range of values and with a measure with 
a normalization that can be fixed later.
This way, we arrive at an expression for the path integral on the finite line of the form
 \begin{equation}
_{out}\la p', \alpha | p, \bar \alpha \ra_{in}\sim 
\int dT \int d \theta\
e^{i s\theta}\
\text{Det}(\partial_\tau -i\theta)\, \text{Det}(\partial_\tau +i \theta)
  \int {DX}\,  e^{iS_{\text{gf}}}\;,
 \end{equation}
where the gauge-fixed action in the exponent reads
 \begin{equation}
S_{\text{gf}}= 
\int_0^1 d\tau\,\Big[-\dot p_\mu x^\mu-i\,\bar\alpha_\mu(\partial_\tau - i \theta) \alpha^\mu- T p^2  \Big] 
-i \bar \alpha(0) \alpha(0)\;.
\label{action-4}
\end{equation}
Here, we have extracted from the action the constant Chern-Simons term, whose coupling $q$ gets renormalized to the physical value $s$. Also, we have
included the Faddeev-Popov determinants, 
and inserted the boundary terms in the action, needed
to match the quantum numbers of the external states.
Performing the path integral, and normalizing appropriately the  integration over the moduli, while setting their correct range, we obtain 
  \ba \label{1.17}
_{out}\la p', \alpha | p, \bar \alpha \ra_{in} &\sim 
\int_0^\infty dT \int_0^{2\pi} \frac{d \theta}{2\pi} \ 
(2\pi)^D \delta^{D}(p-p') \,
e^{is\theta}\, e^{e^{-i\theta } \bar\alpha \alpha}\,  e^{-iTp^2}
\\
&= (2\pi)^D \delta^{D}(p-p') \frac{(-i)}{p^2} \int_0^{2\pi} \frac{d \theta}{2\pi} \, 
 e^{e^{-i\theta } \bar\alpha \alpha + is\theta}\;.
 \ea
Note that the ghost determinant, which has vanishing Dirichlet boundary conditions, becomes $\theta$ independent and can be normalized to 1, see Appendix \ref{Appendix:A} for details.\footnote{The result is similar to that in eq. \eqref{A.15}, but with $\alpha=\bar \alpha= 0$, which selects the vacuum as the external states.}
Finally, performing the $\theta$-integration, we find
  \ba
  _{out}\la p', \alpha | p, \bar \alpha \ra_{in} =
 (2\pi)^D \delta^{D}(p-p') \frac{(-i )}{p^2} \frac{(\bar\alpha \alpha)^s}{s!} 
\;.
 \label{spins}
 \ea
In particular, for spin 1, obtained by setting $s=1$, we get
  \ba
_{out}\la p', \alpha | p, \bar \alpha \ra_{in} =
 (2\pi)^D \delta^{D}(p-p')\, \frac{-i\eta_{\mu\nu}}{p^2}\, \bar\alpha^\mu \alpha^\nu \;.
 \ea
We recognize the propagators of massless fields of spin $s$ in the Feynman gauge. The independent variables $\alpha$ and $\bar \alpha$ parametrize the coherent states and can be readily interpreted as the external polarizations. 

This concludes our discussion of the path integral on the finite line for the free spin-$s$ particle. A natural extension involves introducing a set of ghost fields (henceforth denoted $\cal B$ $\bar {\cal C}$  and $\bar {\cal B}$ $\cal C$) to exponentiate the Faddeev-Popov determinants. Then, considering external states also for them  allows to explore the full BRST Hilbert space.

A final comment concerns the issue of gauge-fixing $u$ and $\bar u$. We observe that these gauge fields can be defined so as not to possess associated moduli,
so that the path integral for the original action \eqref{act-1} can be performed without the need to gauge
the $U(1)$ symmetry. This leads to the propagation of the full set of higher spin fields in the Feynman gauge.
Indeed, the path integral produces
  \ba
 _{out}\la p', \alpha | p, \bar \alpha \ra_{in} =
(2\pi)^D \delta^{D}(p-p') \frac{(-i)}{p^2} \,
 e^{\bar\alpha \alpha}\;,
 \label{allspins}
 \ea
which corresponds precisely to the sum over all spins of Eq. \eqref{spins}.

\subsection{Non-abelian background} \label{sec2.2}
To compute tree-level gluon amplitudes, we couple the bosonic spinning particle to a Yang-Mills background field 
\begin{equation}
\cA_\mu:=A_\mu^a\,T_a\;,
\end{equation}
where $T_a$ are anti-hermitian generators; we will consider color-ordered amplitudes, thus the Lie algebra generators will not play much of a role in the following.
We extend the gauge constraints by minimal coupling $p_\mu\rightarrow p_\mu-i\cA_\mu$, and by adding a non-minimal term with 
coupling constant $\kappa$ to the Hamiltonian,
\ba \label{Covariant constraints}
L_A= \alpha^\mu  (p_\mu-i \cA_\mu) 
\;, \quad \bar L_A=  \bar \alpha^\mu  (p_\mu-i \cA_\mu)\;, \quad H_A= \frac12 (p-i \cA)^2 -\kappa\, \cF_{\mu\nu}\,\alpha^\mu \bar\alpha^\nu\;,
\ea
with $\cF_{\mu\nu}=\del_\mu\cA_\nu-\del_\nu\cA_\mu+[\cA_\mu,\cA_\nu]$.
The constraint algebra \eqref{constraint algebra} is broken by the field strength $\cF_{\mu\nu}$ for any value of $\kappa$, both at the quantum and classical level.
Despite this, suitable spin sectors of the model can be quantized consistently, using BRST techniques; we shall briefly review how such a construction works.

In canonical BRST quantization, one extends the theory by adding fermionic antighost/ghost pairs $(\cB, \bar \cC)$, $(\bar \cB, \cC)$, $(b, c)$
associated with the constraints $(L_A , \bar L_A,  H_A)$, respectively, with anticommutation relations defined by
\be
 \{\mathcal{B},\bar{\mathcal{C}}\} = 1 \;, \quad 
\{\bar{\mathcal{B}},\mathcal{C}\} = 1\;,  \quad 
   \{b, c\} = 1 \;.
\ee 
The quantum BRST operator corresponding to the constraints \eqref{Covariant constraints} is then given by
\begin{equation}\label{QA}
\begin{split}
Q_A&=-2\,c\, H_A+i\,\bar\cC L_A+i\,\cC\bar L_A-\cC\bar\cC\,b\\
&=c\,\big(\cD^\mu\cD_\mu+2\kappa\, \cF_{\mu\nu}\,\alpha^\mu \bar\alpha^\nu\big)+(\bar\cC\alpha^\mu+\cC\bar\alpha^\mu)\cD_\mu-\cC\bar\cC\,b\;,
\end{split}    
\end{equation}
where $\cD_\mu=\del_\mu+\cA_\mu$ and we identified $p_\mu\equiv-i\del_\mu$, keeping for simplicity the same notation for classical and quantum variables. The BRST operator $Q_A$ is hermitian if one assigns suitable hermiticity properties to the ghosts
\be
 c^\dagger\ = c \;, \quad b^\dagger = b \;, \quad \cC^\dagger = -\bar \cC \;, \quad \cB^\dagger = -\bar \cB \;. 
\ee
The extended BRST Hilbert space, on which the BRST charge $Q_A$ operates, has a double grading. 
One corresponds to the usual ghost number. 
The other one is associated with the integer $U(1)$ charge given by the number charge ${\cal N}$, which counts the number of $\alpha^\mu$ oscillators as in \eqref{2.16}, appropriately extended to account for the charge of the ghost oscillators:
\begin{equation}\label{BRST U1}
\cN:=\alpha^\mu\bar\alpha_\mu+\cC\bar\cB+\cB\bar\cC\;.    
\end{equation}
In the previous section, when constructing the free path integral, we gauged the $N$ charge in \eqref{2.16} 
to project on the spin-1 sector. Here, its extended version will still be used as a projector, but we prefer to keep it outside the BRST operator.
It is defined to obey $[Q_A,\cN]=0$, so that it becomes consistent in the following to restrict  the study of the cohomology of the BRST operator 
to a pre-fixed $U(1)$ charge sector.
Quantization is consistent if $Q_A^2=0$. According to the analysis of \cite{Bonezzi:2024emt}, we have the following situation, depending on the spin sector of the theory: 
\begin{itemize}
\item If the background field $\cA_\mu$ is off-shell, only the spin zero sector of the model (i.e., the subspace of the Hilbert space with $\cN=0$) can be quantized consistently. In this case, the coupling $\kappa$ is irrelevant, as the corresponding non-minimal term acts trivially. This sector simply describes a colored scalar field in the adjoint representation of the gauge group.
\item If the background field $\cA_\mu$ is on-shell, i.e. it obeys the nonlinear Yang-Mills equations $\cD^\mu\cF_{\mu\nu}=0$, and for $\kappa=1$ the model can be quantized also in the $\cN=1$ sector. This is the case we are interested in, where the worldline describes a gluon in a Yang-Mills background.
\end{itemize}

In order to extract the gluon vertex operators and study the path integral, we set $\kappa=1$ and proceed with a Hamiltonian BRST gauge-fixing of the classical action.
Upon denoting graded coordinates and momenta by $Z^I:=(x^\mu,\alpha^\mu;c,\bar\cC,\cC)$ and $\cP_I:=(p_\mu,-i\bar\alpha_\mu;-ib,-i\cB,-i\bar\cB)$, respectively, we choose $\Psi=iTb$ as the gauge-fixing fermion related to the gauge choice
$(e,u,\bar u)=(2T,0,0)$. Thus, the gauge-fixed action reads
\begin{equation}\label{coupled-action}
\begin{split}
S&=\int_0^1 d\tau\,\Big[\dot Z^I\cP_I+\big\{Q_A,\Psi\big\}_{\rm PB} 
\Big]\\
&=\int_0^1 d\tau\,\Big[p_\mu\dot x^\mu-i\,\bar\alpha_\mu\dot\alpha^\mu+ib\dot c+i\cB\dot{\bar\cC}+i\bar\cB\dot\cC-2TH_A 
\Big] \;.    
\end{split}    
\end{equation}
Integrating out the momenta, we finally obtain the sigma model action
\begin{equation}\label{Sigma model}
S=S_{\text{free}} +S_{\text{int}}\;,
\end{equation}
where the free Lagrangian action, including the ghosts, reads\footnote{The parameter $\tau$ is understood to be shifted and rescaled.}
\begin{equation} \label{Free action}
S_{\text{free}}=\int_{-T/2}^{T/2}\!\!\!d\tau\Big[\tfrac14\,\dot x^2-i\,\bar\alpha_\mu\dot\alpha^\mu+ib\dot c+i\cB\dot{\bar\cC}+i\bar\cB\dot\cC
\Big] \;,
\end{equation}
while the interaction terms are collected in
\begin{equation} \label{Interacting action}
   S_{\text{int}} =\int_{-T/2}^{T/2}\!\!\!d\tau\Big[i\cA_\mu(x)\dot x^\mu+2\cF_{\mu\nu}(x)\alpha^\mu\bar\alpha^\nu
\Big] \;.
\end{equation}
One can derive vertex operators from the above action in the usual way, upon specializing the background gauge field to a sum of plane waves, with definite momentum and polarization. These will be the so-called
integrated vertex operators, which describe the emission/absorption of a gluon from the worldline interior. One can already see from \eqref{Interacting action} the need for a two-gluon vertex, arising from the coupling to the non-abelian field strength.

In the previous subsection, we have attributed the spin-1 degrees of freedom to the worldline endpoints by employing coherent states for the $\alpha^\mu$ oscillators. In the following, we will find it more convenient to take simpler boundary conditions, corresponding to a suitable vacuum state, and create \emph{all} external gluon states by means of vertex operators. To do so, the vertex operators extracted from the sigma model action \eqref{Interacting action} are not sufficient; therefore, in order to discuss vertex operators more generally, together with the issue of ghost zero modes, we will briefly revert to canonical BRST quantization.

\subsection{Vertex operators from canonical BRST quantization} \label{sec2.3}

Here we will briefly revisit the vertex operators as consistent deformations of the free BRST charge. In particular, we will review how the consistent coupling to the background $\cA_\mu$ translates into the BRST invariance of vertex operators. We will then motivate the definition of integrated vertex operators in the canonical language.

Let us collectively denote bosonic oscillators and ghost variables as
\begin{equation}
\alpha^M:=(\cC,\alpha^\mu,\cB)\;,\quad\bar\alpha_M:=(\bar\cB,\bar\alpha_\mu,\bar\cC)\;,    
\end{equation}
according to their $U(1)$ charge: $[\cN,\alpha^M]=\alpha^M$ and $[\cN,\bar\alpha_M]=-\bar\alpha_M$.
In canonical quantization, we define the BRST ``oscillator vacuum'' $\ket{0}$ as the Fock vacuum annihilated by the operators $\bar\alpha_M$ and by the antighost $b$. In the path integral formulation, the boundary conditions corresponding to this vacuum at both endpoints of the worldline are $\bar\alpha_M(-T/2)=0$, $\alpha^M(T/2)=0$ and $b(\pm T/2)=0$.
Taking also $\dot x^\mu(\pm T/2)=0$, it is natural to view the full BRST vacuum as the tensor product of an eigenstate of zero momentum with $\ket{0}$, i.e.
$\ket{0,0}=\ket{k=0} \otimes\ket{0}$. We thus have
\begin{equation}
\bar\alpha_M\ket{0,0}=0\;,\quad b\ket{0,0}=0\;,\quad p_\mu\ket{0,0}=0\;.   
\end{equation}
Although $\ket{0,0}$ is not normalizable, we will always create states with nonzero momentum by suitable vertex operators at the boundary. These states have the standard plane wave normalization 
\begin{equation}
\bra{k',0}c\ket{k,0}=(2\pi)^D\delta^D(k-k')\;,    
\end{equation}
where we recall that the basic overlap for the $bc$ system is $\bra{0}c\ket{0}=1$.

The subspace of the Hilbert space we are interested in is the $\cN=1$ sector, which describes free spin-1 particles. A generic state in this sector can be written as
\begin{equation}\label{spin one state}
\l x|\psi\r=\psi_M(x)\,\alpha^M\ket{0}+\chi_M(x)\,\alpha^M\,c\ket{0}\;, 
\end{equation}
in the position representation. 
Any such state is annihilated by normal ordered operators with more than one $\bar\alpha_M$, a feature that will be crucial for a consistent coupling to Yang-Mills backgrounds or, equivalently, for the BRST invariance of the vertex operators.

We define the \emph{unintegrated} canonical vertex operator for the gauge field as the linear deformation of the covariant BRST operator \eqref{QA}:
\begin{equation}
Q_A=Q+\cV_A+\cO(A^2)\;,
\end{equation}
where the free (${\cal A}$-independent) BRST differential reads
\begin{equation}\label{Q free}
Q=c\,\B+(\bar\cC\alpha^\mu+\cC\bar\alpha^\mu)\del_\mu-\cC\bar\cC\,b\;,    
\end{equation}
while the linear vertex operator is given by
\begin{equation}
\begin{split}
\cV_A&=(\bar\cC\alpha^\mu+\cC\bar\alpha^\mu)\cA_\mu+c\,\big(2\cA^\mu\del_\mu+\del\cdot\cA+2\,f_{\mu\nu}\,\alpha^\mu \bar\alpha^\nu\big)\;,
\end{split}    
\end{equation}
upon denoting $f_{\mu\nu}=\del_\mu\cA_\nu-\del_\nu\cA_\mu$ the abelian part of the field strength.
Expanding $Q_A^2$ at first order in $\cA_\mu$ produces
\begin{equation}\label{QV}
\{Q,\cV_A\}=c\,(\alpha^\nu\bar\cC-\cC\bar\alpha^\nu)\,\del^\mu f_{\mu\nu}-3\,\cC\alpha^\mu\bar\cC\bar\alpha^\nu\,f_{\mu\nu}-2\,c\,(\cC\alpha^\nu\bar\alpha^\mu\bar\alpha^\rho+\alpha^\mu\alpha^\nu\bar\cC\bar\alpha^\rho)\,\del_\mu f_{\nu\rho}  \;, 
\end{equation}
written in normal ordering. We see that the second and third terms in \eqref{QV} are of the form $\cO^{MN}\bar\alpha_M\bar\alpha_N$, with two annihilation operators on the right. For this reason, they kill any state in the $\cN=1$ sector. Since both $Q$ and $\cV_A$ commute with $\cN$, the restriction to any $\cN=s$ subspace can be explicitly enforced by the projector 
\begin{equation}
\cP_s:=\int_0^{2\pi}\frac{d\theta}{2\pi}\,e^{i\theta(\cN-s)}\equiv\delta_{\cN,s}\;,\quad\cP_s^2=\cP_s\;.    
\end{equation}
Projecting \eqref{QV} to the spin one subspace, we thus obtain
\begin{equation}\label{QV is VQ}
\{Q,\cV_A\}\,\cP_1=c\,(\bar\cC\alpha^\nu-\cC\bar\alpha^\nu)\,\del^\mu f_{\mu\nu}\,\cP_1\;.    
\end{equation}
If the gauge field $\cA_\mu$ obeys the linearized Yang-Mills equations, the unintegrated vertex operator is BRST invariant under projection. It can thus be used to create asymptotic gluon states that satisfy such free equations.

For the path integral on the circle, it is necessary to enforce the constraint by inserting the projector $\cP_1$, which is equivalent to gauging the $U(1)$ symmetry. However, when computing tree-level processes on an open worldline, it is sufficient to have initial and final states with $\cN=1$ to ensure that no state with $\cN\neq1$ is created at any stage. This is so because $\cN$ is conserved in the process, as all vertex operators have charge zero and $\cN$ commutes with the gauge-fixed Hamiltonian. 

In view of computing the path integral on the open line, the above discussion brings a small subtlety. In order to create all external states in the process by inserting vertex operators, the Fock vacuum $\ket{0}$ is not the correct boundary state, since it has charge $\cN=0$. In order to generate physical states with vertex operators, the correct vacuum state is
\begin{equation}
\ket{1}:=\cB\ket{0}\;,\quad\bra{1}:=\bra{0}\bar\cB \;.
\end{equation}
This state has $\cN=1$, obeys $Q\ket{1}=0$ and can be viewed as a constant Yang-Mills ghost. We will later present the boundary conditions corresponding to the physical vacuum $\ket{1}$.

We now come to discuss integrated vertex operators. To this end, we start by splitting the unintegrated vertex $\cV_A$ according to its dependence on the reparametrization ghost $c$:
\begin{equation}
\cV_A=W_A+c\,V_A\;,
\end{equation}
where
\begin{equation}
\begin{split}
W_A&=(\bar\cC\alpha^\mu+\cC\bar\alpha^\mu)\cA_\mu\;,\\  V_A&=2\cA^\mu\del_\mu+\del\cdot\cA+2\,f_{\mu\nu}\,\alpha^\mu \bar\alpha^\nu\;.
\end{split}    
\end{equation}
The (integrand of) the integrated vertex operator is defined via $\{b,\cV_A\}\equiv V_A$. The rationale for this definition is that, if $\{Q,\cV_A\}=0$, the BRST transformation of $V_A$ is given by
\begin{equation}
[Q,V_A]=[\B,\cV_A]\;,    
\end{equation}
upon using $\{Q,b\}=\B$. As we have previously discussed, if $\cA_\mu$ obeys the linearized field equations $\{Q,\cV_A\}$ is not identically zero, but does vanish on the projected Hilbert space.

To integrate the vertex operator $V_A$, we have to introduce time dependence by switching to the Heisenberg picture. Upon rescaling the gauge-fixed Hamiltonian to $p^2=-\B$, we define time-dependent vertex operators via
\begin{equation}
\cV_A(\tau):=e^{iH\tau}\cV_A\,e^{-iH\tau}=e^{ip^2\tau}\cV_A\,e^{-ip^2\tau}\;,    
\end{equation}
with analogous formulas for any other operator. Since $[Q,H]=0$, we can see that the BRST transformation of $V_A(\tau)$ is a total derivative:
\begin{equation}
[Q,V_A(\tau)]=-[p^2,\cV_A(\tau)]=i\,\frac{d}{d\tau}\cV_A(\tau)    \;.
\end{equation}
The BRST variation of the integrated vertex operator is then given by
\begin{equation}
\left[Q,\int_{\tau_1}^{\tau_2} d \tau\,V_A(\tau)\right]=i\,\cV_A(\tau_2)-i\cV_A(\tau_1)\;.    
\end{equation}

Typically, this fails to vanish when the boundaries of the integral reach other insertions. Perhaps, this is the most direct way to see the need for nonlinear vertex operators: they must be added to cure the failure of BRST invariance at these insertion points.

The basic, unintegrated, bilinear vertex can be extracted from the part of $Q_A$ \eqref{QA} quadratic in the gauge field, namely
\begin{equation}\label{V2 BRST}
\cV_{A,A}=c\,\big(\cA^2+2\,[\cA_\mu,\cA_\nu]\alpha^\mu\bar\alpha^\nu\big)  \;.  
\end{equation}
We will discuss the role of this vertex in the computation of amplitudes as we go along.

\subsection{Lagrangian vertex operators and color stripping} \label{sec2.3.1}
In order to compute color-ordered amplitudes, we have to strip off color from the vertex operators and take the gauge field to represent a single gluon state. For both $\cV_A$ and $V_A$, which are linear in $\cA_\mu$, this is simply done by removing the Lie algebra generator and taking the field to be a plane wave with definite momentum and polarization:
\begin{equation}
\cA_\mu(x)=A_\mu^a(x)\,T_a\quad\longrightarrow\quad \epsilon_\mu\,e^{ik\cdot x}\;.    
\end{equation}
The derivative operator $-i\del_\mu$, when appearing in Weyl-ordered expressions, reduces in the path integral approach
to the classical canonical momentum $p_\mu$ without further terms.
For the free action \eqref{Free action} this coincides with $\frac12\,\dot x^\mu$, but in the sigma model \eqref{Sigma model} it is rather given by $\frac12\,\dot x^\mu+i\cA^\mu$. This is ultimately the reason why the nonlinear vertices differ between the Hamiltonian and Lagrangian path integrals. This correspondence allows us to identify
\begin{equation}
2\cA^\mu\del_\mu+\del\cdot\cA\quad\longrightarrow \quad e^{ik\cdot x}\big(2\epsilon^\mu\del_\mu+ik\cdot\epsilon\big)\quad\longrightarrow \quad i\epsilon_\mu\dot x^\mu\,e^{ik\cdot x}\;, 
\end{equation}
where we used the fact that the operator expression $2\,\cA^\mu\del_\mu+\del\cdot\cA$ is already Weyl ordered.\footnote{Weyl ordering is tied to the construction of a regulated, background gauge-invariant, path integral, see e.g. \cite{Bastianelli:2006rx}.}
Proceeding in the same way for all terms in $\cV_A$ and $V_A$, we find the one-gluon color-stripped vertex operator
\begin{equation}\label{Vertex operators}
\cV_{\epsilon,k}(\tau)=W_{\epsilon,k}(\tau)+c(\tau)\,V_{\epsilon,k}(\tau)\;,
\end{equation}
where we recall that $V_{\epsilon,k}(\tau)$ is to be used for integrated vertices:
\begin{equation}
\begin{split} \label{Vertex operator V}
W_{\epsilon,k}(\tau)&=\epsilon_\mu\Big(\cC(\tau)\,\bar\alpha^\mu(\tau)+\bar\cC(\tau)\,\alpha^\mu(\tau)\Big)\,e^{ik\cdot x(\tau)}\;,\\
V_{\epsilon,k}(\tau)&=\Big(i\,\epsilon_\mu\,\dot x^\mu(\tau)+2i\,f_{\mu\nu}(k)\,\alpha^\mu(\tau)\,\bar\alpha^\nu(\tau)\Big)\,e^{ik\cdot x(\tau)}\;,
\end{split}
\end{equation}
and we defined $f_{\mu\nu}(k)=k_\mu\epsilon_\nu-k_\nu\epsilon_\mu$.

The above vertex operators are not enough to compute arbitrary amplitudes. 
As already mentioned, further nonlinear vertex operators are needed. The simplest one is bilinear and can be seen directly from the action or the BRST charge. In its Hamiltonian form, this bilinear vertex is extracted from the BRST charge \eqref{QA} and is given by $\cV_{A,A}$ in  \eqref{V2 BRST}.
To properly strip off color, we assign the two gluon states respecting the order in which the generators are removed:
\begin{equation}
\begin{split}
\cA^2+2\,[\cA_\mu,\cA_\nu]\alpha^\mu\bar\alpha^\nu&=\big(A^a\cdot A^b+4\,A^a_\mu A^b_\nu\,\alpha^{[\mu}\bar\alpha^{\nu]}\big)T_aT_b\\
&\longrightarrow\quad\big(\epsilon_1\cdot\epsilon_2+4\,\epsilon_{1\mu} \epsilon_{2\nu}\,\alpha^{[\mu}\bar\alpha^{\nu]}\big)\,e^{i(k_1+k_2)\cdot x}\;.
\end{split}    
\end{equation}
The bilinear operator corresponding to the second term
\begin{equation} \label{bl}
V_{12}(\tau)=4\,\epsilon_{1\mu} \epsilon_{2\nu}\,\alpha^{[\mu}(\tau)\bar\alpha^{\nu]}(\tau)\,e^{i(k_1+k_2)\cdot x(\tau)}  \;,  
\end{equation}
where we have used the shorthand notation $V_{ij}=V_{\epsilon_i,k_i;\epsilon_j,k_j}$,
can be clearly derived from the action \eqref{Interacting action}. The first term corresponding to $\cA^2$, however, is absent from the Lagrangian action. This is because the canonical momentum is given by $\frac12\,\dot x^\mu+i\cA^\mu$. The effects of the bilinear vertex $\epsilon_1\cdot\epsilon_2\,e^{i(k_1+k_2)\cdot x(\tau)}$ are correctly taken into account by contractions of the $\dot x^\mu$ factors with each other in products $V_{\epsilon_1,k_1}(\tau)\,V_{\epsilon_2,k_2}(\sigma)$. The vertex operator \eqref{bl} is thus the only two-gluon vertex to be used in the Lagrangian path integral, as indeed implied by the interacting part of the sigma model action \eqref{Interacting action}.

Let us comment on the color structure of the bilinear vertex. In computing full amplitudes, including color, the bilinear vertex operator for particles $i$ and $j$ contributes a color factor of the form ${\rm tr}(\cdots T_iT_j\cdots)$. This implies that the two gluons created by $V_{ij}$ are necessarily \emph{color-adjacent}, which means that, in computing color-ordered amplitudes, $V_{ij}$ is needed only if gluons $i$ and $j$ are adjacent in the cyclic ordering $12\ldots n$.

Finally, as we shall discuss, additional nonlinear vertex operators are required for the construction of amplitudes. While the linear and quadratic vertex operators are already encoded in the action -- or equivalently, in the BRST charge -- further composite vertex operators, which attach subtrees to a main worldline, arise from the requirement of BRST-invariant amplitudes, thereby ensuring the consistency of the theory. Although a derivation of these operators from purely worldline considerations remains perhaps incomplete, the analysis presented in \cite{Bonezzi:2025iza} for the scalar theory contains the essential ingredients. We will return to this issue in the next sections.

\section{Tree-level gluon amplitudes} \label{sec3}
We now apply the formalism developed in the previous sections to compute gluon scattering amplitudes at tree level. To this end, following the analysis of \cite{Bonezzi:2025iza}, we take the worldline to have infinite length, so that the proper time $T$ serves purely as an infrared regulator and drops out of all computations. The amplitudes are obtained by evaluating suitable correlation functions with insertions of the vertex operators, previously derived by coupling the worldline to a Yang-Mills background field. 
Before delving into the explicit calculations, we briefly outline how correlators are defined and evaluated on open worldlines in our formalism. In what follows, all external gluons are taken to be on-shell, though not necessarily transverse.

\subsection{Correlation functions on open lines} \label{sec3.1}
To project on states of zero momentum, the coordinate trajectories are split as
\begin{equation}\label{BC Neumann z}
x^\mu(\tau)=x^\mu_0+z^\mu(\tau)\;,\quad \dot z^\mu(\pm T/2)=0\;, 
\end{equation}
where $x_0^\mu$ is the zero mode, which enforces momentum conservation. Coming to the oscillator vacuum, recall that the boundary conditions corresponding to $\ket{0}$ at both $\tau=\pm T/2$ are
\begin{equation}
\bra{0}\ldots\ket{0}\quad\longleftrightarrow \quad\bar\alpha_M(-T/2)=0\;,\quad\alpha^M(T/2)=0\;,\quad b(\pm T/2)=0\;.   
\end{equation}
On the other hand, the physical vacuum $\ket{1}=\cB\ket{0}$ is characterized by $(\cB,b,\bar\cB)\ket{1}=0$. Upon denoting all ghosts and antighosts as $C^i:=(\cC,c,\bar\cC)$ and $B_i:=(\bar\cB,b,\cB)$, the boundary conditions corresponding to the vacuum $\ket{1}$ are then
\begin{equation}\label{BC for amplitudes}
\bra{1}\ldots\ket{1}\quad\longleftrightarrow \quad\bar\alpha_\mu(-T/2)=0\;,\quad\alpha^\mu(T/2)=0\;,\quad B_i(\pm T/2)=0\;.   
\end{equation}
Since all ghosts and antighosts have the same boundary conditions, we can rewrite the free Lagrangian action \eqref{Free action} as
\begin{equation}\label{Free action for amplitudes}
S_{\text{free}}=\int_{-T/2}^{T/2}\!\!\!d\tau\Big[\tfrac14\,\dot x^2-i\,\bar\alpha_\mu\dot\alpha^\mu+iB_i\dot C^i
\Big] \;.        
\end{equation}
Note that the basic overlap for the physical vacuum needs three ghost insertions:
\begin{equation}
\bra{1}\cC\,c\,\bar\cC\ket{1}=1\;.
\end{equation}
In the path integral, this is reflected by the fact that all $C^i(\tau)$ admit a zero mode, which needs to be saturated. The basic expectation value for the ghosts is thus given by
\begin{equation}\label{Ghost saturation}
\l\cC(\tau_1)\,c(\tau_2)\,\bar\cC(\tau_3)\r=1\;.   
\end{equation}
From the quadratic action \eqref{Free action for amplitudes} with boundary conditions \eqref{BC Neumann z} and \eqref{BC for amplitudes}, we derive the following two-point functions (the ghost zero mode saturation is left implicit)
\begin{equation}\label{z-z}
\begin{split}
\l\bar\alpha^\mu(\tau)\alpha^\nu(\sigma)\r&=\eta^{\mu\nu}\theta(\tau-\sigma)\;,\\
\l z^\mu(\tau)z^\nu(\sigma)\r&=-i\eta^{\mu\nu}G(\tau,\sigma)\;,\\
\l \dot z^\mu(\tau)z^\nu(\sigma)\r&=-i\eta^{\mu\nu}\udot G(\tau,\sigma)\;,\\
\l \dot z^\mu(\tau)\dot z^\nu(\sigma)\r&=-i\eta^{\mu\nu}\udot G\udot (\tau,\sigma)\;,
\end{split}    
\end{equation}
where we denote derivatives with respect to the left and right variables by left and right dots, respectively. The $z$-propagator and its derivatives read
\begin{equation}\label{z-z-1}
\begin{split}
G(\tau,\sigma)&=|\tau-\sigma|-\frac{1}{T}\,(\tau-\sigma)^2-\frac{2}{T}\,\tau\sigma-\frac{T}{6}\;,\\
\udot G(\tau,\sigma)&=\epsilon(\tau-\sigma)-\frac{2}{T}\,\tau\;,\\
\udot G\udot(\tau,\sigma)&=-2\,\delta(\tau-\sigma)\;.
\end{split}    
\end{equation}
In particular, notice that the Neumann boundary condition $\udot G(\pm T/2,\sigma)=0$ holds only at separate points. At coincident points one rather has $\udot G(\pm T/2,\pm T/2)=\mp1$. This reflects the Weyl ordering of operators, that is implicit in the path integral.\footnote{ For instance, the quantum operator corresponding to $x^\mu p^\nu$ is $\frac12\,(\hat x^\mu\hat p^\nu+\hat p^\nu\hat x^\mu)$.} This fact will be relevant when computing the three-gluon amplitude.

\subsection{Color-ordered amplitudes} \label{sec3.2}
We now illustrate the construction above with explicit computations, starting from the simplest test of the model: the three-gluon amplitude. Then, we move on to the calculation of the color-ordered four-point amplitude, which we perform in two distinct ways. First, we select the worldline connecting two of the least color-adjacent particles, which we regard as a ``smart" choice for computational reasons, as should be clear shortly. Then, we reproduce the same calculation by choosing the ``not-so-smart" worldline, namely the one that connects two color-adjacent particles. Throughout this section, we implicitly factor out the momentum-conserving delta function, which arises from the integral over the zero-mode $x_0^\mu$. Every $x^\mu(\tau)$ in the vertex operators can thus be substituted with the fluctuation $z^\mu(\tau)$.

\subsubsection{The three-gluon amplitude} \label{sec3.2.1}
Although the three-gluon amplitude vanishes for real momenta due to kinematical reasons, its expression is still nontrivial and can be compared with the standard field theory result. Recall that we take external gluons to be on-shell but not necessarily transverse, so that $k_i^2=0$, $k_i\cdot\epsilon_i\neq0$. 

To saturate the ghost zero modes, we need all three vertex operators to be unintegrated. This has a clear geometric interpretation: two vertex operators are fixed at asymptotic times $\tau=\pm\infty$ to create the incoming and outgoing worldline states, while the third one is fixed at an intermediate time using the rigid translation invariance on the infinite line. To compute the color-ordered amplitude, we need to insert the three vertex operators with a fixed (cyclic) order in time (see figure \ref{fig1}). The remaining expectation value is taken with respect to $z^\mu$ and the other variables and gives the color-ordered amplitude as
\begin{equation}
\cA_{123}=\big\l\cV_1(+\infty)\cV_2(0)\cV_3(-\infty)\big\r\;,  
\end{equation}
where we used the shorthand $i=\{\epsilon_i,k_i\}$ for particle labels. 

    \diagramThree

Using the split $\cV_i(\tau_i)=W_i(\tau_i)+c(\tau_i)\,V_i(\tau_i)$, the above expression gives three contributions
(as one needs only one $c$-ghost insertion for obtaining a nonvanishing result) 
\begin{equation}\label{A3 split}
\begin{split}
\cA_{123}&=\big\l W_1(+\infty)\,c(0)V_2(0)W_3(-\infty)\big\r+\big\l W_1(+\infty)W_2(0)\,c(-\infty)V_3(-\infty)\big\r\\    
&\phantom{=}+\big\l c(+\infty)V_1(+\infty)W_2(0)W_3(-\infty)\big\r \;.    
\end{split}
\end{equation}
The expression \eqref{Vertex operator V} for $V_i(\tau_i)$ suggests that $V_i(\pm\infty)=0$, thanks to the boundary conditions. As we have already mentioned, however, this is not true due to the self-contraction
\begin{equation}
\epsilon\cdot\dot z(\tau)\,e^{ik\cdot z(\tau)}\sim k\cdot\epsilon\udot G(\tau,\tau)\,e^{ik\cdot z(\tau)}+\cdots\;,  
\end{equation}
and $\udot G(\tau,\tau)$ does not vanish when $\tau\rightarrow\pm\infty$. This term does vanish for transverse gluons, which motivates us to consider non-transverse polarizations in order to study its effect.

Let us compute separately the three contributions in \eqref{A3 split}. All correlators have a common ``Koba-Nielsen''
factor, given by the contraction of the plane waves $e^{ik_i\cdot z(\tau_i)}$, which is trivial, due to momentum conservation and the on-shell condition:
\begin{equation}
\begin{split}
\exp\Big\{\tfrac{i}{2}\sum_{ij}k_i\cdot k_j\,G(\tau_i,\tau_j)\Big\}&=\exp\Big\{i\sum_{i<j}k_i\cdot k_j\,|\tau_i-\tau_j|\Big\}\\
&=\exp\left\{-\frac{iT(k_1^2+k_3^2)}{2}\right\}=1\;,
\end{split}
\end{equation}
where we used $T/2\rightarrow\infty$ as an infrared regulator. We further use \eqref{Ghost saturation} for saturating the ghost zero modes. This only gives relative signs between the three contributions. The first term in \eqref{A3 split} yields
\begin{equation}
\begin{split}
&\big\l W_1(+\infty)\,c(0)V_2(0)W_3(-\infty)\big\r\\
&=\Big\l\epsilon_1\cdot\bar\alpha(+\infty)\,\big[i\dot z(0)\cdot\epsilon_2+4i\,k_2^{[\mu}\epsilon_2^{\nu]}\alpha_\mu(0)\bar\alpha_\nu(0)\big]\epsilon_3\cdot\alpha(-\infty)e^{i\sum_ik_i\cdot z(\tau_i)}\Big\r\\
&=\epsilon_1\cdot\epsilon_3\,i\epsilon_2\cdot\sum_ik_i\udot G(0,\tau_i)+4i\,k_2^{[\mu}\epsilon_2^{\nu]}\epsilon_{1\mu}\epsilon_{3\nu}\\
&=i\Big(\epsilon_1\cdot\epsilon_3\,\epsilon_2\cdot(k_3-k_1)+2\,\epsilon_1\cdot k_2\,\epsilon_3\cdot\epsilon_2-2\,\epsilon_1\cdot \epsilon_2\,\epsilon_3\cdot k_2\Big)\;,
\end{split}    
\end{equation}
where in the second line we removed the ghost factor. The above result is already the correct amplitude for transverse gluons. The remaining two terms add the contributions from the longitudinal modes:
\begin{equation}
\begin{split}
&\big\l W_1(+\infty)W_2(0)\,c(-\infty)V_3(-\infty)\big\r\\
&=-\Big\l\epsilon_1\cdot\bar\alpha(+\infty)\,\epsilon_2\cdot\alpha(0)\,i\epsilon_3\cdot\dot z(-\infty)\,e^{i\sum_ik_i\cdot z(\tau_i)}\Big\r\\
&=-\epsilon_1\cdot\epsilon_2\,ik_3\cdot\epsilon_3\udot G(-\infty,-\infty)=-i\,\epsilon_1\cdot\epsilon_2\,k_3\cdot\epsilon_3\;,
\end{split}    
\end{equation}
where we used the fact that only the term with $\dot z^\mu$ in $V_i(\pm\infty)$ contributes. The last term in \eqref{A3 split} similarly gives
\begin{equation}
\begin{split}
&\big\l c(+\infty)V_1(+\infty)W_2(0)W_3(-\infty)\big\r\\
&=-\Big\l i\epsilon_1\cdot\dot z(+\infty)\,\epsilon_2\cdot\bar\alpha(0)\,\epsilon_3\cdot\alpha(-\infty)\,\,e^{i\sum_ik_i\cdot z(\tau_i)}\Big\r\\
&=-\epsilon_2\cdot\epsilon_3\,ik_1\cdot\epsilon_1\udot G(+\infty,+\infty)=i\epsilon_2\cdot\epsilon_3\,k_1\cdot\epsilon_1\;.
\end{split}    
\end{equation}
By using momentum conservation, we can write the full result in the manifest cyclic form
\begin{equation}\label{A123}
\begin{split}
\cA_{123}&=
i\,\Big[\epsilon_1\cdot\epsilon_2\,\epsilon_3\cdot(k_1-k_2)+\epsilon_2\cdot\epsilon_3\,\epsilon_1\cdot(k_2-k_3)+\epsilon_3\cdot\epsilon_1\,\epsilon_2\cdot(k_3-k_1)\Big]\;.
\end{split}    
\end{equation}
This is the correct three-point amplitude even for non-transverse gluons, as it coincides with the color-ordered cubic vertex seen in the second quantized picture.
\subsubsection{The four-gluon amplitude}  \label{sec3.2.2}
For the four-point color-ordered amplitude, we start by selecting the ``smart" worldline, i.e. the one connecting particles 4 in the past with particle 2 in the future (or, equivalently, the one connecting particles 1 and 3, see figure \ref{fig2}).\footnote{It seems that the most convenient choice is a worldline that connects two of the least color-adjacent particles.} The advantage in this choice of worldline comes from the fact that we are allowed to integrate $V_1(\tau)$ over the entire line, without altering the cyclic order of the particles. We are thus led to evaluate
\begin{equation}
\cA_{1234}=
\Big\l\cV_2(+\infty)\, \cV_3(0) \, \int_{-\infty}^{+\infty} {\hskip -6mm d\tau}\, V_1(\tau)\, \cV_4(-\infty)
\Big\r\;,  
\end{equation}
which, for transverse gluons ($V_{\epsilon,k}(\pm\infty)=0$), reduces to 
\begin{equation}
\cA_{1234}=
\Big\l W_2(+\infty)\, c(0)\,V_3(0) \, \int_{-\infty}^{+\infty} {\hskip -6mm d\tau}\, V_1(\tau)\, W_4(-\infty)
\Big\r\;.
\end{equation}
Considering the overlap $\l \cC\, c\ \bar\cC\r =1 $, it simplifies to 
\begin{equation}
\cA_{1234}=
\Big\l 
\epsilon_2\cdot\bar\alpha(+\infty) e^{i k_2\cdot z(+\infty)}\, 
 V_3(0) \, \int_{-\infty}^{+\infty} {\hskip -6mm d\tau}\, V_1(\tau)\, 
 \epsilon_4\cdot \alpha(-\infty) e^{i k_4\cdot z(-\infty)}
 \Big\r\;.
\end{equation} 
The overall exponentials of the vertex operators will lead by Wick contractions to 
\begin{equation}
\begin{split}
\big\l 
e^{i \sum_i  k_i\cdot z(\tau_i)}  
 \big\r\
&=
e^{-\frac12 \sum_{ij}  k_i\cdot \l z(\tau_i)  z(\tau_j) \r \cdot k_j} 
 =
  e^{\frac{i}{2}  \sum_{ij}  k_i \cdot k_j G(\tau_i,\tau_j)}
\\
&=
 e^{i  \sum_{i<j}  k_i \cdot k_j |\tau_i -\tau_j|}
 =
 e^{\frac{i}{2} (- s_{12} \tau + s_{23} \tau + s_{31} |\tau|) }\;,
\end{split}
\end{equation} 
 where we have used the basic two-point functions in \eqref{z-z} and \eqref{z-z-1},
 noted that only the absolute value in $G(\tau_i,\tau_j)$ contributes, 
 and used the Mandelstam variables 
\begin{equation}
s_{ij}= (k_i+k_j)^2 = 2 k_i \cdot k_j \;.
\end{equation}
To display the separation in two channels, the final answer is better written using the on-shell value of $s_{31}=-s_{12}-s_{23}$, leading to 
\begin{equation}\label{Bosonic core}
\big\l 
e^{i \sum_i  k_i\cdot z(\tau_i)}  
 \big\r\
 = e^{\theta(\tau)(-i \tau s_{12}) +\theta(-\tau)(i \tau s_{23})} =
 \theta(\tau)\, e^{ -i \tau s_{12}} +
 \theta(-\tau)\, e^{  i \tau s_{23}} \;.
\end{equation} 
The integral in $\tau$ will split into the two regions $\tau>0$ and $\tau<0$, corresponding to the $s$- and $t$-channels, respectively (recall that there is no $u$-channel in the color-ordered amplitude).

    \diagramFourSmart

The remainder of the correlation function yields the numerator contributions. To compute these, we can further split the vertex operators $V_i(\tau)$ into their ``scalar'' and ``spin'' parts as
\begin{equation} \label{split}
V_{i}(\tau)=V^{\rm scal}_{i}(\tau)+V^{\rm spin}_{i}(\tau)\;
\end{equation}
with
\begin{equation}
\begin{split}
V^{\rm scal}_{i}(\tau)&=i\,\epsilon_i\cdot\dot z(\tau)\,e^{ik_i\cdot z(\tau)}\;,\\
V^{\rm spin}_{i}(\tau)&=4i\,k_i^{[\mu}\epsilon_i^{\nu]}\alpha_\mu(\tau)\,\bar\alpha_\nu(\tau)\,e^{ik_i\cdot z(\tau)}\;.  
\end{split}    
\end{equation}
Omitting the contraction \eqref{Bosonic core} of the plane waves, these terms contribute to the correlator as follows:
\begin{equation}
\begin{split} \label{4.50}
&\big\l 
\cdots\,V^{\rm scal}_3(0)\, V^{\rm scal}_1(\tau)\cdots\big\r=4\,\theta(\tau)\,\epsilon_2\cdot\epsilon_4\,\epsilon_1\cdot k_2\,\epsilon_3\cdot k_4+4\,\theta(-\tau)\,(1\leftrightarrow3)-2i\,\delta(\tau)\,\epsilon_2\cdot\epsilon_4\,\epsilon_1\cdot\epsilon_3\;,\\[3mm] 
&\big\l 
\cdots\,V^{\rm scal}_3(0)\, V^{\rm spin}_1(\tau)\cdots
 \big\r+\big\l 
\cdots\,V^{\rm spin}_3(0)\, V^{\rm scal}_1(\tau)\cdots\big\r=\\
&4\,\theta(\tau)\,\Big[\big(\epsilon_2\cdot k_3\,\epsilon_3\cdot\epsilon_4-\epsilon_2\cdot \epsilon_3\,k_3\cdot\epsilon_4\big)\,\epsilon_1\cdot k_2-\big(\epsilon_2\cdot k_1\,\epsilon_1\cdot\epsilon_4-\epsilon_2\cdot \epsilon_1\,k_1\cdot\epsilon_4\big)\,\epsilon_3\cdot k_4\Big]+4\,\theta(-\tau)\,(1\leftrightarrow3)\;,\\[3mm]
&\big\l 
\cdots\,V^{\rm spin}_3(0)\, V^{\rm spin}_1(\tau)\cdots\big\r=\\
&4\,\theta(\tau)\,\Big[\epsilon_4\cdot k_3\big(\epsilon_3\cdot\epsilon_1\,k_1\cdot\epsilon_2-\epsilon_3\cdot k_1\,\epsilon_1\cdot\epsilon_2\big)-\epsilon_4\cdot \epsilon_3\big(k_3\cdot\epsilon_1\,k_1\cdot\epsilon_2-k_3\cdot k_1\,\epsilon_1\cdot\epsilon_2\big)\Big]+4\,\theta(-\tau)\,(1\leftrightarrow3)\;,
\end{split}    
\end{equation}
where $+\,(1\leftrightarrow3)$ means to exchange simultaneously $(\epsilon_1,k_1)$ with $(\epsilon_3,k_3)$. Summing the three contributions above, multiplied by the common scalar factor \eqref{Bosonic core}, and performing the $\tau$ integral gives the final result. This can be organized upon introducing the following products:
\begin{equation}
\begin{split} \label{C-inf prod}
m_2^\mu(\epsilon_i,\epsilon_j)&:=i\,\Big(2\,\epsilon_i\cdot k_j\,\epsilon_j^\mu-2\,\epsilon_j\cdot k_i\,\epsilon_i^\mu+(k^\mu_i-k^\mu_j)\,\epsilon_i\cdot\epsilon_j\Big)\;,\\
m_3^\mu(\epsilon_i,\epsilon_j,\epsilon_k)&:=\epsilon_i\cdot\epsilon_j\,\epsilon_k^\mu+\epsilon_k\cdot\epsilon_j\,\epsilon_i^\mu-2\,\epsilon_i\cdot\epsilon_k\,\epsilon_j^\mu\;.
\end{split}    
\end{equation}
These products encode (for transverse gluons) the color-ordered Feynman rules for the cubic and quartic vertices, respectively \cite{Zeitlin:2008cc, Bonezzi:2022yuh, Bonezzi:2023xhn}, and play a central role in the Berends-Giele recursion relations \cite{Berends:1987me, Selivanov:1997aq, Mizera:2018jbh, Lopez-Arcos:2019hvg}. In terms of the products above, one can rewrite the result coming from the path integral as
\begin{equation}
\cA_{1234}=\frac{i}{s_{12}}\,m_2(\epsilon_1,\epsilon_2)\cdot m_2(\epsilon_3,\epsilon_4)+\frac{i}{s_{23}}\,m_2(\epsilon_2,\epsilon_3)\cdot m_2(\epsilon_4,\epsilon_1)+i\,\epsilon_2\cdot m_3(\epsilon_3,\epsilon_4,\epsilon_1)\;,    
\end{equation}
which highlights the two exchange diagrams and the contact quartic term. This is the correct color-ordered four-gluon amplitude. It is customary to blow up the quartic vertex and define kinematic numerators as
\begin{equation}
\begin{split}
n_{12}&:=m_2(\epsilon_1,\epsilon_2)\cdot m_2(\epsilon_3,\epsilon_4)+\frac{s_{12}}{3}\,\big[m_3(\epsilon_1,\epsilon_2,\epsilon_3)-m_3(\epsilon_2,\epsilon_1,\epsilon_3)\big]\cdot\epsilon_4\;,\\
n_{23}&:=m_2(\epsilon_2,\epsilon_3)\cdot m_2(\epsilon_1,\epsilon_4)+\frac{s_{23}}{3}\,\big[m_3(\epsilon_2,\epsilon_3,\epsilon_1)-m_3(\epsilon_3,\epsilon_2,\epsilon_1)\big]\cdot\epsilon_4\;,\\
n_{31}&:=m_2(\epsilon_3,\epsilon_1)\cdot m_2(\epsilon_2,\epsilon_4)+\frac{s_{31}}{3}\,\big[m_3(\epsilon_3,\epsilon_1,\epsilon_2)-m_3(\epsilon_1,\epsilon_3,\epsilon_2)\big]\cdot\epsilon_4\;,
\end{split}    
\end{equation}
where the various channels are sometimes identified with the standard names for the Mandelstam variables for four points: $12\rightarrow s$, $23\rightarrow t$, $31\rightarrow u$. Upon using the following symmetry properties of the products:
\begin{equation}
\begin{split}
m_2(\epsilon_i,\epsilon_j)+m_2(\epsilon_j,\epsilon_i)&=0\;,\\
m_3(\epsilon_i,\epsilon_j,\epsilon_k)-m_3(\epsilon_k,\epsilon_j,\epsilon_i)&=0\;,\\
m_3(\epsilon_i,\epsilon_j,\epsilon_k)+m_3(\epsilon_j,\epsilon_k,\epsilon_i)+m_3(\epsilon_k,\epsilon_i,\epsilon_j)&=0\;,
\end{split}    
\end{equation}
the color-ordered amplitude can be recast in the compact form
\begin{equation}
\cA_{1234}=i\,\left[\frac{n_s}{s}-\frac{n_t}{t}\right] \;.
\end{equation}

We shall now illustrate how a different choice for the worldline would lead to the same result, albeit with comparatively greater effort.
\paragraph{Not-so-smart worldline}
Choosing instead the worldline connecting the color-adjacent particles 1 and 4 in figure \ref{fig2}, we have to compute three different contributions
\begin{equation}
\cA_{1234}=\cA_{1234}^{\mathrm{I}}+\cA_{1234}^{\mathrm{II}}+\cA_{1234}^{\mathrm{III}}\;,
\end{equation}
corresponding to three different diagrams (see figure \ref{fig3}). Explicitly
\begin{align}
    \cA_{1234}^{\mathrm{I}}&=-\Big\l\cV_1(+\infty) \, \int_{0}^{+\infty} {\hskip -6mm d\tau}\, V_2(\tau) \, \cV_3(0) \, \cV_4(-\infty)\Big\r\ , \label{AI}\\
    \cA_{1234}^{\mathrm{II}}&= i\,\Big\l\cV_1(+\infty) \, c(0) V_{23}(0)\, \cV_4(-\infty)
\Big\r\ , \label{AII} \\
    \cA_{1234}^{\mathrm{III}}&= i\,\Big\l\cV_1(+\infty) \, c(0)\,V_{23}^{\rm pinch}(0)\, \cV_4(-\infty) \Big\r\; , \label{AIII}
\end{align}
where the phases are given by a factor of $i$ for every insertion corresponding to a real interaction vertex. The first contribution $\cA_{1234}^{\mathrm{I}}$ arises from integrating the linear vertex for particle 2, with the range restricted to preserve the cyclic ordering of particles. Since gluons 2 and 3 are color-adjacent, the bilinear vertex $V_{23}(\tau)$ is also needed, as discussed in sec.~\ref{sec2.3.1}. Using translation invariance, we place it at $\tau=0$. 
These two contributions, however, are not sufficient to produce a gauge invariant amplitude, as we will demonstrate in sec.~\ref{sec4}. The missing part, which we denoted $\cA_{1234}^{\mathrm{III}}$ above, involves a composite operator, $V_{23}^{\rm pinch}(\tau)$, which attaches to the worldline the cubic subtree formed by particles 2 and 3. This pinch operator is nothing but a vertex operator $V_{\epsilon_{23},k_{23}}$ of the standard form \eqref{Vertex operator V}, with momentum $k_2+k_3$ and multiparticle polarization given by $\epsilon^\mu_{23}=\tfrac{1}{s_{23}} m_2^\mu(\epsilon_2,\epsilon_3)$, to be further discussed in the next section. 
    \diagramFourDumb

We start the computation with the first diagram $\cA_{1234}^{\mathrm{I}}$ \eqref{AI}, whose calculation is akin to the one for the smart worldline. Taking transverse gluons and recalling the overlap $\l \cC\, c\ \bar\cC\r =1$, it simplifies to
\begin{equation} \label{4.60}
\cA_{1234}^{\mathrm{I}}=-\big\l 
\epsilon_1\cdot\bar\alpha(+\infty) e^{i k_1\cdot z(+\infty)} \, \int_{0}^{+\infty} {\hskip -6mm d\tau}\, V_2(\tau)\, V_3(0) \,
 \epsilon_4\cdot \alpha(-\infty) e^{i k_4\cdot z(-\infty)}
 \big\r\;,
\end{equation} 
with the Koba-Nielsen scalar factor reducing to 
\begin{equation} \label{fact}
\big\l 
e^{i \sum_i  k_i\cdot z(\tau_i)}  
 \big\r\
= e^{-i \, s_{12} \tau}\;.
\end{equation} 
Splitting the vertex operators into scalar and spin parts as in \eqref{split}, and omitting the contraction of the plane waves, the numerator contributions are given by the following correlators
\begin{equation}
\begin{split}
&\big\l 
\cdots\,V^{\rm scal}_2(\tau)\,V^{\rm scal}_3(0)\,\cdots\big\r=-2i \, \delta(\tau) \, \epsilon_1 \cdot \epsilon_4 \, \epsilon_2 \cdot \epsilon_3 +4 \,  \epsilon_1 \cdot \epsilon_4 \, \epsilon_2 \cdot k_1 \, \epsilon_3 \cdot k_4\;,\\[3mm] 
&\big\l 
\cdots\,V^{\rm spin}_2(\tau)\,V^{\rm scal}_3(0)\,\cdots
 \big\r+\big\l 
\cdots\,V^{\rm scal}_2(\tau)\,V^{\rm spin}_3(0)\,\cdots\big\r=\\
&-4\left[ \epsilon_3 \cdot k_4 \left( \epsilon_1 \cdot k_2 \, \epsilon_2 \cdot \epsilon_4 - \epsilon_1 \cdot \epsilon_2 \, k_2 \cdot \epsilon_4 \right) - \epsilon_2 \cdot k_1 \left( \epsilon_1 \cdot k_3 \, \epsilon_3 \cdot \epsilon_4 -\epsilon_1 \cdot \epsilon_3 \, k_3 \cdot \epsilon_4 \right)\right]\;,\\[3mm]
&\big\l 
\cdots\,V^{\rm spin}_2(\tau)\,V^{\rm spin}_3(0)\,\cdots\big\r=\\
&-4\left[ k_3 \cdot \epsilon_4 \left( \epsilon_1 \cdot \epsilon_2 \, k_2 \cdot\epsilon_3 - \epsilon_2 \cdot \epsilon_3 \, \epsilon_1 \cdot k_2  \right) - \epsilon_3 \cdot \epsilon_4 \left( \epsilon_1 \cdot \epsilon_2 \, k_2 \cdot k_3 -\epsilon_1 \cdot k_2 \, \epsilon_2 \cdot k_3 \right) \right]\;,
\end{split}    
\end{equation}
which is the same as in \eqref{4.50} but for $\tau>0$ and with $(1\leftrightarrow2)$. Multiplying by the common scalar factor \eqref{fact} and integrating over $\tau$ we get
\begin{align}\nonumber
&\big\l 
\cdots\, V^{\rm scal}_2(\tau)\,V^{\rm scal}_3(0)\,\cdots\big\r=i \epsilon_1 \cdot \epsilon_4 \, \left(\epsilon_2 \cdot \epsilon_3+\frac{4 \epsilon_2 \cdot k_1 \, \epsilon_3 \cdot k_4}{s_{12}}\right)\;,\\[3mm]\nonumber 
&\big\l 
\cdots\,V^{\rm spin}_2(\tau)\,V^{\rm scal}_3(0)\,\cdots
 \big\r+\big\l 
\cdots\,V^{\rm scal}_2(\tau)\,V^{\rm spin}_3(0)\,\cdots\big\r=\\\nonumber
&-\frac{4i}{s_{12}}\left[ \epsilon_3 \cdot k_4 \left( \epsilon_1 \cdot k_2 \, \epsilon_2 \cdot \epsilon_4 - \epsilon_1 \cdot \epsilon_2 \, k_2 \cdot \epsilon_4 \right) - \epsilon_2 \cdot k_1 \left( \epsilon_1 \cdot k_3 \, \epsilon_3 \cdot \epsilon_4 -\epsilon_1 \cdot \epsilon_3 \, k_3 \cdot \epsilon_4 \right) \right]\;,\\[3mm]\nonumber
&\big\l 
\cdots\,V^{\rm spin}_2(\tau)\,V^{\rm spin}_3(0)\,\cdots\big\r=\\
&-\frac{4 i}{s_{12}}\left[  k_3 \cdot \epsilon_4 \left( \epsilon_1 \cdot \epsilon_2 \, k_2 \cdot\epsilon_3 - \epsilon_2 \cdot \epsilon_3 \, \epsilon_1 \cdot k_2  \right) - \epsilon_3 \cdot \epsilon_4 \left( \epsilon_1 \cdot \epsilon_2 \, k_2 \cdot k_3 -\epsilon_1 \cdot k_2 \, \epsilon_2 \cdot k_3 \right) \right]\;.
\end{align}
The second diagram \eqref{AII}, with the bilinear vertex 
\begin{equation}
V_{23}(0)=\Big(2\,\epsilon_{2}\cdot\alpha(0)\,\epsilon_3\cdot\bar\alpha(0)-2\,\epsilon_{3}\cdot\alpha(0)\,\epsilon_2\cdot\bar\alpha(0)\Big)\,e^{i(k_2+k_3)\cdot x(0)}  \ ,  
\end{equation}
yields the local contribution
\begin{equation}
\cA_{1234}^{\mathrm{II}}=2i \left( \epsilon_1 \cdot \epsilon_2 \, \epsilon_3 \cdot \epsilon_4-\epsilon_1 \cdot \epsilon_3 \, \epsilon_2 \cdot \epsilon_4 \right)\ .
\end{equation}
We can recast the following partial contribution in a clearer form through the products \eqref{C-inf prod}: 
\begin{equation}
    \cA_{1234}^{\mathrm{I}}+\cA_{1234}^{\mathrm{II}}=\frac{i}{s_{12}} \, m_{2}(\epsilon_1 ,\epsilon_2) \cdot m_{2}(\epsilon_3 ,\epsilon_4) +i\,\epsilon_2\cdot m_3(\epsilon_3,\epsilon_4,\epsilon_1)\ . 
\end{equation}
Finally, the third contribution \eqref{AIII} can be computed with the following pinch operator
\begin{equation}
  V_{23}^{\rm pinch}(0)=\frac{i}{s_{23}} m_{2 \,\mu}(\epsilon_2 ,\epsilon_3) \left(\dot{x}^\mu(0) +4 (k_2+k_3)_\nu \, \alpha^{[\nu}(0)\bar{\alpha}^{\mu]} (0) \, e^{i(k_2+k_3)\cdot x(0)} \right)\;.
\end{equation}
Since the scalar factor coming from the plane waves is $1$, this contribution reduces to
\begin{equation}
\begin{split}
&\left.\big\l 
\cdots\,V_{23}^{\rm pinch}(0)\,\cdots\big\r\right|_{\rm scal}=\frac{i}{s_{23}} \, \epsilon_1 \cdot \epsilon_4 \,  m_{2}(\epsilon_2 ,\epsilon_3) \cdot (k_4-k_1)\;,\\[3mm] 
&\left.\big\l 
\cdots\,V_{23}^{\rm pinch}(0)\,\cdots\big\r\right|_{\rm spin}=\frac{2i}{s_{23}} m_{2 \, \mu}(\epsilon_2 ,\epsilon_3) \left[ \epsilon_1 \cdot (k_2+k_3) \, \epsilon_4^\mu -\epsilon_4 \cdot (k_2+k_3) \, \epsilon_1^\mu  \right] \;, 
\end{split}    
\end{equation}
which yields the $t$-channel part of the amplitude
\begin{equation}
\begin{split}
\cA_{1234}^{\mathrm{III}}&=-\frac{1}{s_{23}} \, m_{2 \, \mu}(\epsilon_2 ,\epsilon_3) \left[\epsilon_1 \cdot \epsilon_4 \, (k_4-k_1)^\mu +2\epsilon_1 \cdot (k_2+k_3) \, \epsilon_4^\mu -2\epsilon_4 \cdot (k_2+k_3) \, \epsilon_1^\mu  \right]\\
&=\frac{i}{s_{23}} \, m_{2}(\epsilon_2 ,\epsilon_3) \cdot m_{2}(\epsilon_4 ,\epsilon_1)\;.
\end{split}    
\end{equation}
The final result is thus given by the sum of the three individual contributions, reproducing the expected expression
\begin{equation}
\cA_{1234}=\frac{i}{s_{12}}\,m_2(\epsilon_1,\epsilon_2)\cdot m_2(\epsilon_3,\epsilon_4)+\frac{i}{s_{23}}\,m_2(\epsilon_2,\epsilon_3)\cdot m_2(\epsilon_4,\epsilon_1)+i\,\epsilon_2\cdot m_3(\epsilon_3,\epsilon_4,\epsilon_1)\;.  
\end{equation}
This shows that the worldline connecting two color-adjacent gluons requires as many correlators as field theory Feynman diagrams. In addition, we have verified that the outcome does not depend on the choice of the main worldline.

\subsection{Path integral representation for Berends-Giele currents}  \label{sec3.3}
As just seen in the latter case, the pinching contribution, where gluons 2 and 3 are fused into a cubic subtree, requires the introduction of the composite field
\begin{equation}\label{A23}
A_{23}^\mu(x)=\frac{1}{k_{23}^2}\,m_2^\mu(\epsilon_2,\epsilon_3)\,e^{ik_{23}\cdot x}  \;.  
\end{equation}
The form of $A_{23}^\mu$ is usually determined by the Bern-Kosower pinching rules, or from the field theory color-ordered Feynman rules. In this last part, we will show that the composite field can be computed directly from a worldline path integral, with different boundary conditions.

In order to describe nonlinear fields via worldline correlators, we follow the scalar example studied in \cite{Bonezzi:2025iza}, and consider a semi-infinite worldline. We take the affine parameter $\tau\in(-\infty,0]$, so that the worldline has one real boundary and an asymptotic one. Since the proper length of the line is still infinite, there is no modulus for the einbein, which we gauge fix to $e(\tau)=2$. At the asymptotic past we want, as before, to project onto the zero-momentum vacuum $\ket{1}=\cB\ket{0}$, so that we can create an asymptotic gluon state by inserting $\cV_{\epsilon,k}(-\infty)$. The vacuum $\ket{1}$ corresponds to the asymptotic boundary conditions
\begin{equation}
\dot x^\mu(-\infty)=0\;,\quad\bar\alpha^\mu(-\infty)=0\;,\quad\bar\cB(-\infty)=\cB(-\infty)=b(-\infty)=0 \;.   
\end{equation}
At $\tau=0$ instead, we want to project onto a position eigenstate which, in order to support a vector field, is at the same time a coherent state for $\alpha^\mu$: $\bra{x,\alpha}c$. This bra state corresponds to the boundary conditions
\begin{equation}
x^\mu(0)=x^\mu\;,\quad\alpha^\mu(0)=\alpha^\mu\;,\quad \cB(0)=\cC(0)=c(0)=0\;.    
\end{equation}
Note that the appearance of the $c$ ghost in the bra state $\bra{x,\alpha}c$, corresponding to the boundary condition $c(0)=0$, is due to the absence of $c$ zero-modes on the semi-infinite line.  
The BRST invariant action with these boundary conditions is given by
\begin{equation}
S_{\rm DN}=\int_{-\infty}^0\!\!\!d\tau\,\Big[\tfrac14\,\dot x^2-i   \,\bar\alpha\cdot\dot\alpha+i\,b\dot c+i\,\bar\cB\dot\cC+i\,\bar\cC\dot\cB\Big]\;,    
\end{equation}
where the subscript stands for Dirichlet-Neumann. 

Coherently with the boundary conditions, the only ghost that admits a zero mode is $\bar\cC$, so that the saturation requires $\big\l\bar\cC(\tau)\big\r=1$. To compute correlation functions, we expand the fields as backgrounds plus fluctuations
\begin{equation}
x^\mu(\tau)=x^\mu+z^\mu(\tau)\;,\quad\alpha^\mu(\tau)=\alpha^\mu+\kappa^\mu(\tau)\;,\quad\bar\alpha^\mu(\tau)=\bar\kappa^\mu(\tau)\;,    
\end{equation}
with vanishing (asymptotic) boundary conditions
\begin{equation}
z^\mu(0)=0\;,\quad\dot z^\mu(-\infty)=0\;,\quad\kappa^\mu(0)=0\;,\quad\bar\kappa^\mu(-\infty)=0\;.    
\end{equation}
These lead to the two-point functions
\begin{equation}
\begin{split}
\l z^\mu(\tau)\,z^\nu(\sigma)\r&=-i\,\eta^{\mu\nu}\,G_{\rm DN}(\tau,\sigma)\;,\quad \l \bar\kappa^\mu(\tau)\,\kappa^\nu(\sigma)\r=\eta^{\mu\nu}\,\theta(\tau-\sigma)\;,\\
G_{\rm DN}(\tau,\sigma)&=|\tau-\sigma|+(\tau+\sigma)\;,\quad \udot G_{\rm DN}(\tau,\sigma)=\epsilon(\tau-\sigma)+1\;,
\end{split}
\end{equation}
with the Dirichlet-Neumann propagator obeying $G_{\rm DN}(0,\sigma)=0$ and $\udot G_{\rm DN}(-\infty,\sigma)=0$.

As a first example, we compute the one-point function of an unintegrated vertex operator placed at $\tau=-T$, to be sent to infinity. The path integral over the fluctuations is normalized to one, and we denote correlation functions by $\l\cdots\r_{\rm DN}$ to remind us of the nontrivial boundary conditions. The correlator yields
\begin{equation}
\begin{split}
\Big\l\cV_{\epsilon,k}(-T)\Big\r_{\rm DN}&=e^{ik\cdot x}\Big\l\epsilon\cdot\big(\alpha+\kappa(-T)\big)\,e^{ik\cdot z(-T)}\bar\cC(-T)\Big\r_{\rm DN}\\
&=e^{ik\cdot x}\epsilon\cdot \alpha\Big\l e^{ik\cdot z(-T)}\bar\cC(-T)\Big\r_{\rm DN}=\alpha^\mu\epsilon_\mu\,e^{ik\cdot x}e^{-iTk^2}\;.
\end{split}    
\end{equation}
We see that, as for the infinite line, if the momentum of the asymptotic state is off-shell, the correlator vanishes in the limit $T\rightarrow\infty$, while for $k^2=0$ it gives $\alpha^\mu\epsilon_\mu\,e^{ik\cdot x}$, thus reproducing a linear on-shell state.

As a more interesting application, we try to reproduce the nonlinear field $A_{23}^\mu$ in \eqref{A23} by inserting $\cV_3(-\infty)$ and integrating $V_2(\tau)$ over the line. Since both describe asymptotic particles, the momenta are on-shell and their polarizations are transverse. Taking care of the ghost zero mode saturation, the correlator reads
\begin{equation}
\begin{split}
i\int_{-\infty}^0\!\!\!d\tau\,\Big\l V_2(\tau)\cV_{3}(-\infty)\Big\r_{\rm DN}&=i\,e^{ik_{23}\cdot x}\Big\l \Big(i\epsilon_2\cdot\dot z(\tau)+4ik_2^{[\mu}\epsilon_2^{\nu]}\big(\alpha_\mu+\kappa_\mu(\tau)\big)\bar\kappa_\nu(\tau)\Big)e^{ik_2\cdot z(\tau)}\\
&\phantom{=}\times\epsilon_3\cdot\big(\alpha+\kappa(-\infty)\big)e^{ik_3\cdot z(-\infty)}\bar\cC(-\infty)\Big\r_{\rm DN}   \;.
\end{split}    
\end{equation}
The contraction of the two plane waves yields the common factor
\begin{equation}
\big\l e^{ik_2\cdot z(\tau)}e^{ik_3\cdot z(-\infty)}\big\r_{z}=e^{ik_{23}^2\tau}\;,
\end{equation}
where the subscript refers to the path integral over $z$. The term proportional to $\alpha\cdot\epsilon_3$ contributes as
\begin{equation}
\begin{split}
i\alpha\cdot\epsilon_3\Big\l e^{ik_2\cdot z(\tau)}e^{ik_3\cdot z(-\infty)}i\epsilon_2\cdot\dot z(\tau)\Big\r_{z}&=-\alpha\cdot\epsilon_3\,\epsilon_2\cdot k_3\,e^{ik_{23}^2\tau}\udot G_{\rm DN}(\tau,-\infty)\\
&=-2\,\alpha\cdot\epsilon_3\,\epsilon_2\cdot k_3\,e^{ik_{23}^2\tau}\;,
\end{split}    
\end{equation}
while the term with $\epsilon_3\cdot\kappa(-\infty)$ yields
\begin{equation}
\begin{split}
-4  \,k_2^{[\mu}\epsilon_2^{\nu]}\Big\l e^{ik_2\cdot z(\tau)}e^{ik_3\cdot z(-\infty)}\big(\alpha_\mu+\kappa_\mu(\tau)\big)\bar\kappa_\nu(\tau)\,\epsilon_3\cdot\kappa(-\infty)\Big\r_{z,\kappa}=-4\,k_2^{[\mu}\epsilon_2^{\nu]}\,\alpha_\mu\epsilon_{3\nu}\,e^{ik_{23}^2\tau}\;.
\end{split}    
\end{equation}
Finally, integrating over $\tau$ we obtain
\begin{equation}
\begin{split} \label{BGtilde}
i\int_{-\infty}^0\!\!\!d\tau\,\Big\l V_2(\tau)\cV_{3}(-\infty)\Big\r_{\rm DN}&=\frac{2i}{k_{23}^2}\,\alpha_\mu\Big(\epsilon_2\cdot k_3\,\epsilon_3^\mu-\epsilon_3\cdot k_2\,\epsilon_2^\mu+k_2^\mu\epsilon_2\cdot\epsilon_3\Big)\,e^{ik_{23}\cdot x}\\&=\frac{1}{k_{23}^2}\,\alpha_\mu\Big(m_2^\mu(\epsilon_2,\epsilon_3)+i\,k_{23}^\mu\,\epsilon_2\cdot\epsilon_3\Big)\,e^{ik_{23}\cdot x}\\
&=\alpha_\mu\,\Big(A_{23}^\mu(x)+\del^\mu\lambda_{23}(x)\Big)=\alpha_\mu\,\tilde A_{23}^\mu(x)\;,
\end{split}    
\end{equation}
with the composite gauge parameter
\begin{equation}\label{composite lambda}
\lambda_{23}(x)=\frac{1}{k_{23}^2}\,\epsilon_2\cdot\epsilon_3\,e^{ik_{23}\cdot x}\;.   
\end{equation}
We have thus shown that the path integral with Dirichlet-Neumann boundary conditions reproduces the desired field $A_{23}^\mu$, modulo a linear gauge transformation.
We conclude by noting that, in the computation of the four-gluon amplitude, the difference between $\tilde A_{23}^\mu$ and $A_{23}^\mu$ drops from the correlation function, thus resulting in the same amplitude. 

\section{Ward identities}  \label{sec4}
To establish that the amplitudes computed with this method are gauge invariant, we have to discuss the BRST symmetry of the path integral. Let us recall that to compute amplitudes we use the free action
\begin{equation}
S_{\text{free}}=\int_{-\infty}^{+\infty}\!\!\!d\tau\Big[\tfrac14\,\dot x^2-i\,\bar\alpha_\mu\dot\alpha^\mu+iB_i\dot C^i
\Big] \;,        
\end{equation}
with boundary conditions
\begin{equation}
\dot x^\mu(\pm\infty)=0\;,\quad \alpha^\mu(+\infty)=0\;,\quad\bar\alpha^\mu(-\infty)=0\;,\quad B_i(\pm\infty)=0\;.   
\end{equation}
The above action with these boundary conditions is invariant under the BRST transformations
\begin{equation}\label{BRST transf lagrangian}
\begin{split}
sx^\mu&=c\,\dot x^\mu-i(\cC\bar\alpha^\mu+\bar\cC\alpha^\mu)\;,\\
s\alpha^\mu&=\tfrac12\,\cC\dot x^\mu\;,\quad s\bar\alpha^\mu=-\tfrac12\,\bar\cC\dot x^\mu\;,\\
sc&=i\,\cC\bar\cC\;,\quad s\cC=0\;,\quad s\bar\cC=0\;,\\
sb&=\tfrac{i}{4}\,\dot x^2\;,\quad s\cB=\tfrac12\,\alpha\cdot\dot x-i\,\cC b\;,\quad s\bar\cB=\tfrac12\,\bar\alpha\cdot\dot x+i\,\bar\cC b\;.
\end{split}    
\end{equation} These transformations can be modified by adding trivial terms that vanish on-shell as, for instance, $c\dot c$ which can be added to $sc$. In \eqref{BRST transf lagrangian} we chose the simplest form, with no such terms. The BRST differential so defined is nilpotent only on-shell: $s^2\approx0$, where by $\approx$ we denote equality up to equations of motion. This is expected from the gauge-fixed theory in Lagrangian form. If one reintroduces momenta $p_\mu$, the differential $s$ can be made nilpotent off-shell. The Noether charge corresponding to the BRST symmetry is given by
\begin{equation}
Q=-\tfrac14\,c\,\dot x^2+\tfrac{i}{2}\,(\cC\bar\alpha+\bar\cC\alpha)\cdot\dot x-\cC\bar\cC\, b\;,    
\end{equation}
which yields the free BRST operator \eqref{Q free} upon canonical quantization. In the path integral formulation, one can use the charge $Q$ inside correlation functions to generate BRST transformations. More specifically, given a functional $F$, its BRST transformation is given by the equal time commutator
\begin{equation}
[Q,F](\tau)=i\big(sF\big)(\tau)\;.    
\end{equation}

\subsection{OPE and equal-time commutators} \label{sec4.1}
To discuss equal time commutators and other operator relations in the path integral formulation, we start with a brief detour into the operator product expansion (OPE). Similar to string theory, we will call operator product the product of functions, viewed inside correlators in the path integral:
\begin{equation}
F(\tau)\,G(\sigma):=\big\l\cdots F(\tau)\,G(\sigma)\cdots\big\r\;. 
\end{equation}
These are then computed using Wick contractions and the basic two-point functions. The simplest example is the OPE of two $z^\mu$ fluctuations, yielding
\begin{equation}
z^\mu(\tau)\,z^\nu(\sigma)=-i\,\eta^{\mu\nu}\,G(\tau,\sigma)+:z^\mu(\tau)\,z^\nu(\sigma):\;.    
\end{equation}
The normal ordering symbol amounts to the prescription of not performing Wick contractions within it. This immediately leads to $\l\,:\!F(z)\!:\,\r=F(0)$, where $F(z)$ is an analytic function of $z^\mu$. Another simple example is the OPE of $\dot z^\mu$ (related on shell to the momentum operator via $p^\mu=\frac12\,\dot z^\mu$) with a normal-ordered function of $z$:
\begin{equation}\label{zdot F OPE}
\dot z^\mu(\tau)\,:F\big(z(\sigma)\big):\;=-i\,\left(\epsilon(\tau-\sigma)-\frac{2\tau}{T}\right):\del^\mu F\big(z(\sigma)\big):+ :\dot z^\mu(\tau)F\big(z(\sigma)\big):\;,    
\end{equation}
where we used the two-point function $\udot G(\tau,\sigma)=\epsilon(\tau-\sigma)-2\tau/T$.
The OPE can then be used to define equal time commutators via the regularized prescription
\begin{equation}\label{Equal time commutator}
\begin{split}
[A,B](\tau)&=\lim_{\epsilon\rightarrow0}\Big(A(\tau+\epsilon)-A(\tau-\epsilon)\Big)\,B(\tau)=\lim_{\epsilon\rightarrow0}\int_{\tau-\epsilon}^{\tau+\epsilon}\!\!\!\!\!\!d\sigma\;\dot A(\sigma)\,B(\tau)\\
&=\lim_{\epsilon\rightarrow0}A(\tau)\,\Big(B(\tau-\epsilon)-B(\tau+\epsilon)\Big)=-\lim_{\epsilon\rightarrow0}\int_{\tau-\epsilon}^{\tau+\epsilon}\!\!\!\!\!\!d\sigma\; A(\tau)\,\dot B(\sigma)\;,
\end{split}    
\end{equation}
where the equivalence of the two definitions complies with the (graded) antisymmetry of the commutator.
An important point of the above definition is that one takes the limit $\epsilon\rightarrow0$ \emph{after} computing the OPE at separate points. In particular, this assumes that the \emph{only} operators in the interval $[\tau-\epsilon,\tau+\epsilon]$ are $A$ and $B$, for otherwise the result would not yield just the commutator $[A,B](\tau)$. This is an important subtlety that will play a role in the following. 

As an example, we can compute the commutator $[\dot z^\mu,:\!F(z)\!:]$. Using \eqref{zdot F OPE} and the definition for the commutator we obtain
\begin{equation}
[\dot z^\mu,:\!F(z)\!:](\tau)=-2i\,:\del^\mu F\big(z(\tau)\big):+\lim_{\epsilon\rightarrow0}\Big(:\dot z^\mu(\tau+\epsilon)F\big(z(\tau)\big):-:\dot z^\mu(\tau-\epsilon)F\big(z(\tau)\big):\Big)\;.    
\end{equation}
Naively, $\dot z^\mu(\tau+\epsilon)=\dot z^\mu(\tau)$, since $\ddot z^\mu(\tau)\approx0$ on-shell. Inside correlation functions, $\ddot z^\mu(\tau)$ contributes only to contact terms, since $\ddot z(\tau)\,z(\sigma)\sim\delta(\tau-\sigma)$. Now, it is important that in the regularized definition \eqref{Equal time commutator} one assumes that no other operator sits in the interval $[\tau-\epsilon,\tau+\epsilon]$. This ensures that $\lim_{\epsilon
\rightarrow0}:\dot z^\mu(\tau+\epsilon)F\big(z(\tau)\big):$ is regular, finally yielding
\begin{equation}
[\dot z^\mu,:\!F(z)\!:](\tau)=-2i\,:\del^\mu F\big(z(\tau)\big): \;,   
\end{equation}
which is the expected result. One can similarly define the operator product at coincident points via Weyl ordering:
\begin{equation}
(AB)(\tau):=\lim_{\epsilon\rightarrow0}\,\frac12\,\Big(A(\tau+\epsilon)+A(\tau-\epsilon)\Big)\,B(\tau)\;,    
\end{equation}
which reproduces the OPE computed with the two-point functions at coincident points. For instance, taking the coincidence limit of \eqref{zdot F OPE}, one has
\begin{equation}\label{Weyl ordered dot z F}
\dot z^\mu(\tau)\,:F\big(z(\tau)\big):\;=\frac{2i\tau}{T}\,:\del^\mu F\big(z(\tau)\big):+ :\dot z^\mu(\tau)F\big(z(\tau)\big):\;.   
\end{equation}

A second useful example is the commutator of $\dot z^2$ with functions of $z$, since $\frac14\,\dot z^2$ corresponds to the gauge-fixed Hamiltonian. Let us mention that one has to use the normal ordered expression $:\dot z^2(\tau):$ to avoid divergences from self-contractions. Keeping this in mind, we omit explicit normal ordering symbols in the initial functions. To compute the commutator, we evaluate the OPE at separate points:
\begin{equation}
\frac14\,\dot z^2(\tau\pm\epsilon)\,F\big(z(\tau)\big)=-\frac14\,\left(\pm1-\frac{2\tau}{T}\right)^2:\B F(z):-\frac{i}{2}\,\left(\pm1-\frac{2\tau}{T}\right)\,:\dot z^\mu\del_\mu F(z):+\frac14\,:\dot z^2F(z):    
\end{equation}
where on the right-hand side all $z^\mu$ are evaluated at $\tau$. Above, we have already taken the limit $\epsilon\rightarrow0$, taking into account that it is regular inside normal ordered expressions. The commutator is readily evaluated as
\begin{equation}
\tfrac14\,[\dot z^2,F(z)]=\frac{2\tau}{T}\,:\B F(z):-i:\dot z^\mu\del_\mu F(z):\;,    
\end{equation}
where, again, all $z^\mu$ are $z^\mu(\tau)$. The above expression may look somewhat unfamiliar because of the $\B F$ contribution linear in time. However, looking at the result \eqref{Weyl ordered dot z F} one can see that this is exactly the Weyl ordered expression at coincident points, so that
\begin{equation}
\tfrac14\,[\dot z^2,F(z)]=-i\,\dot z^\mu\del_\mu F(z)=-i\,\frac{dF}{d\tau}\;,    
\end{equation}
which is the Heisenberg equation of motion.

\subsection{BRST cohomology of vertex operators} \label{sec4.2}
We can now show that on-shell vertex operators with transverse gluons are elements of the BRST cohomology. We start by showing that vertex operators with polarization $\epsilon_\mu=ik_\mu$ are trivial, manifesting the on-shell residual gauge symmetry. Recall the unintegrated vertex operator $\cV_{\epsilon,k}(\tau)$:
\begin{equation}
\begin{split}
\cV_{\epsilon,k}(\tau)&=W_{\epsilon,k}(\tau)+c(\tau)\,V_{\epsilon,k}(\tau)\;,\\
W_{\epsilon,k}(\tau)&=\epsilon_\mu\Big(\cC(\tau)\,\bar\alpha^\mu(\tau)+\bar\cC(\tau)\,\alpha^\mu(\tau)\Big)\,e^{ik\cdot x(\tau)}\;,\\
V_{\epsilon,k}(\tau)&=i\,\epsilon_\mu\Big(\dot x^\mu(\tau)+2\,k\cdot\alpha(\tau)\,\bar\alpha^\mu(\tau)-2\,k\cdot\bar\alpha(\tau)\,\alpha^\mu(\tau)\Big)\,e^{ik\cdot x(\tau)}\;.
\end{split}
\end{equation}
Taking the polarization to be proportional to momentum, we have
\begin{equation}
\cV_{ik,k}(\tau)=i\big(\cC\bar\alpha+\bar\cC\alpha\big)\cdot k\,e^{ik\cdot z}-c\,k\cdot\dot z\,e^{ik\cdot z}\;,    
\end{equation}
where we have removed the zero mode $x_0$, which in amplitudes only gives momentum conservation. Using the OPE as described in the previous section, we see that $\cV_{ik,k}(\tau)$ is BRST-exact: 
\begin{equation}
\cV_{ik,k}(\tau)=[Q,e^{ik\cdot z}](\tau) \;.   
\end{equation}
In a similar fashion, the integrated vertex operator $V_{ik,k}(\tau)$ is a total derivative:
\begin{equation}
V_{ik,k}(\tau)=i\,\frac{d}{d\tau}\left(e^{ik\cdot z(\tau)}\right)\;.    
\end{equation}

Coming now to BRST closure, we compute the $Q$-commutator of $\cV_{\epsilon,k}$. Using the on-shell condition and transversality, we obtain
\begin{equation}\label{QV OPE}
\{Q,\cV_{\epsilon,k}\}=-3\,:f_{\mu\nu}(z)\,\alpha^\mu\bar\alpha^\nu\cC\bar\cC:-2i\,:c\,(\cC\bar\alpha+\bar\cC\alpha)\cdot k\,f_{\mu\nu}(z)\,\alpha^\mu\bar\alpha^\nu:\;,    
\end{equation}
where $f_{\mu\nu}(z)=2i\,k_{[\mu}\epsilon_{\nu]}e^{ik\cdot z}$. This is nothing but the functional version of \eqref{QV}. We will now show that the right-hand side above vanishes inside all correlation functions. In particular, we want to prove that
\begin{equation}\label{Normal order vanishing}
\big\l\cdots:F(\tau)\bar\alpha^\mu(\tau)\bar\cC(\tau):\cdots\big\r=0\;,\quad\l\cdots:F(\tau)\bar\alpha^\mu(\tau)\bar\alpha^\nu(\tau):\cdots\r=0\;,  
\end{equation}
for any functional $F(\tau)$ and with arbitrary insertions of vertex operators outside the normal ordering. This is the path integral version of the statement, discussed in sec. \ref{sec2.3}, that normal ordered operators with more than one barred oscillator annihilate every state in the $\cN=1$ sector of the Hilbert space.

To prove the first relation, we first notice that the equation of motion $\del_\tau{\bar\cC}=0$ holds everywhere on the line, since there are no insertions of the conjugate antighost $\cB$. We can then push the ghost to the infinite past $\bar\cC(\tau)=\bar\cC(-\infty)$, where it saturates its zero mode. Now we write $\bar\alpha^\mu(\tau)$ as
\begin{equation}
\bar\alpha^\mu(\tau)=\int_{-\infty}^\tau d\sigma\,\del_\sigma\bar\alpha^\mu(\sigma)\;,    
\end{equation}
using the boundary condition $\bar\alpha^\mu(-\infty)=0$. The integral above is zero almost everywhere, since the equation of motion $\del_\tau\bar\alpha^\mu=0$ holds except at insertions of $\alpha^\nu$, where it gives a contraction with a delta function. At this stage, we need to invoke the $U(1)$ invariance of the vertex operators: every insertion of $\alpha^\nu$ necessarily comes with either a ghost $\bar\cC$, or another $\bar\alpha^\rho$. In the first case, the insertion does not contribute, since there is already a factor of $\bar\cC$. In the second case, the above integral produces a factor of $\bar\alpha^\rho$ at the insertion point. One then starts the same procedure with $\bar\alpha^\rho$ until it reaches the asymptotic past, which concludes the proof. An analogous reasoning can be used to prove the second equation in \eqref{Normal order vanishing}.

Since \eqref{Normal order vanishing} holds in all correlation functions, we use it as an operator equation in the sense of OPE.  
We have thus shown that, inside correlation functions, 
 the unintegrated vertex operators are $Q$-closed:
\begin{equation}
\{Q,\cV_{\epsilon,k}\}=0\;.    
\end{equation}
Using the same arguments, in particular that equations of motion hold when regularizing commutators by point splitting, one shows that the $Q$-commutator of the integrated vertex operator is a total derivative:
\begin{equation}\label{Q Vintegrated}
[Q,V_{\epsilon,k}]=i\,\frac{d}{d\tau}\,\cV_{\epsilon,k}\;.    
\end{equation}
In the following, we will show how to use these relations to prove Ward identities.

\subsection{Ward identities from BRST invariance} \label{sec4.3}
To see how to prove Ward identities, we start from the somewhat trivial example of the three-point amplitude $\cA_{123}$.
For transverse gluons, the amplitude \eqref{A123}
\begin{equation}
\cA_{123}=i\,\Big[\epsilon_1\cdot\epsilon_2\,\epsilon_3\cdot(k_1-k_2)+\epsilon_2\cdot\epsilon_3\,\epsilon_1\cdot(k_2-k_3)+\epsilon_3\cdot\epsilon_1\,\epsilon_2\cdot(k_3-k_1)\Big]\;,
\end{equation}
vanishes when taking any polarization $\epsilon^\mu_i$ to be proportional to the corresponding momentum $k^\mu_i$, which is the Ward identity. In the worldline formulation, the amplitude is given by the correlator
\begin{equation}
\cA_{123}=\big\l\cV_1(+\infty)\,\cV_2(0)\,\cV_3(-\infty)\big\r\;.    
\end{equation}
If we take the particle 1 to have $\epsilon^\mu_1=ik^\mu_1$, the corresponding vertex operator is exact: $\cV_1(\tau)=[Q,e^{ik_1\cdot z}](\tau)$. The asymptotic vacuum $\ket{1}$ is BRST invariant, which translates to the boundary condition $Q(\pm\infty)=0$. Since the vertex operator $\cV_1$ is inserted at $T/2\rightarrow+\infty$, the equal time commutator reduces to
\begin{equation}
\cV_1(T/2)=-\lim_{\epsilon\rightarrow0}e^{ik_1\cdot z(T/2)}Q(T/2-\epsilon)\;.    
\end{equation}

The idea is to ``move'' the BRST charge to the asymptotic past, where it annihilates the vacuum. To do so, we use the fact that $Q$ is conserved, so that $\dot Q=0$ except at insertions of other operators. Since there is no operator insertion until $\tau=0$, we have $Q(T/2-\epsilon)=Q(\tau_*)$ for any $\tau_*\in(0,T/2-\epsilon]$. To move $Q$ across $\cV_2(0)$ we pick a commutator contribution at $\tau=0$. After that, $Q(\tau)$ remains constant for negative times until it approaches $\cV_3(-T/2)$: 
\begin{equation}
Q(T/2-\epsilon)\,\cV_2(0)=-\cV_2(0)\,Q(-T/2+\epsilon)+\{Q,\cV_2\}(0)=-\cV_2(0)\,Q(-T/2+\epsilon)\;,    
\end{equation}
upon using the closure of $\cV_2(0)$. Since $\cV_3$ is at the boundary, and $Q(-\infty)=0$, approaching it from the left coincides with the commutator:
\begin{equation}
\lim_{\epsilon\rightarrow0}Q(-T/2+\epsilon)\,\cV_3(-T/2)=\{Q,\cV_3\}(-T/2)=0\;,
\end{equation} 
which proves the Ward identity $\cA_{k_123}=0$.

For the next example, we consider the four-point amplitude, computed with the smart worldline:
\begin{equation}
\cA_{1234}=\int_{-\infty}^{+\infty}\!\!\!\!\!\!d\tau \,\Big\l\cV_2(+\infty)\,\cV_3(0)\,V_1(\tau)\,\cV_4(-\infty)\Big\r\;.   
\end{equation}
Upon taking $\epsilon_2^\mu=ik_2^\mu$ we have again that the corresponding vertex operator is $Q$-exact:
\begin{equation}
\cV_2(T/2)=-\lim_{\epsilon\rightarrow0}e^{ik_2\cdot z(T/2)}Q(T/2-\epsilon)\;.    
\end{equation}
We can use the same arguments to push $Q$ to the asymptotic past, except when it has to go through the integrated vertex $V_1(\tau)$, picking a contribution from the commutator. Upon regularizing the asymptotic time with $T/2\rightarrow+\infty$ we obtain
\begin{equation}
\begin{split}
-\lim_{\epsilon\rightarrow0}\,&\Big\l e^{ik_2\cdot z(T/2)}Q(T/2-\epsilon) \,\cV_3(0)\,V_1(\tau)\,\cV_4(-T/2)\Big\r\\
&=\Big\l e^{ik_2\cdot z(T/2)}\cV_3(0)\,[Q,V_1](\tau)\,\cV_4(-T/2)\Big\r=i\,\Big\l e^{ik_2\cdot z(T/2)}\cV_3(0)\,\dot\cV_1(\tau)\,\cV_4(-T/2)\Big\r \;,  
\end{split}    
\end{equation}
where we used \eqref{Q Vintegrated}. Upon integrating over $\tau$, we end up with
\begin{equation}
\cA_{1k_234}=\lim_{T\rightarrow\infty}i\,\Big\l e^{ik_2\cdot z(T/2)}\cV_3(0)\,\big(\cV_1(T/2)-\cV_1(-T/2)\big)\,\cV_4(-T/2)\Big\r   
\end{equation}
where the notation $\cA_{1k_234}$ highlights the use of a longitudinal polarization for particle 2.
Regardless of the detailed structure, the above expression vanishes in the limit $T\rightarrow\infty$, because it has off-shell plane waves at asymptotic times. To see this, let us focus on the first term above with $\cV_1(T/2)$. The Wick contraction of the plane waves gives
\begin{equation}
\big\l e^{i(k_1+k_2)\cdot z(T/2)}e^{ik_3\cdot z(0)}e^{ik_4\cdot z(-T/2)} \big\r=\exp\left\{-\tfrac{i}{2}\,(k_1+k_2)^2T\right\}\xrightarrow{T\rightarrow\infty}0\;,    
\end{equation}
since $(k_1+k_2)^2=s\neq0$. We thus see that the Ward identity $\cA_{1k_234}=0$ holds, thanks to the unrestricted integration range of $V_1(\tau)$. From this discussion, one can also see that off-shell correlation functions (for which $T$ is a modulus) do not obey linear Ward identities.

\subsubsection{Ward identity for the color-adjacent worldline}
Using the not-so-smart worldline connecting particles 1 and 4, the four-gluon color-ordered amplitude requires three separate correlation functions. Recalling that every vertex (except the ones at infinity) requires an additional factor of $i$, these are given by \eqref{AI}, \eqref{AII} and \eqref{AIII}
\begin{align}
    \cA_{1234}^{\mathrm{I}}&=(i)^2\Big\l\cV_1(+\infty) \, \int_{0}^{+\infty} {\hskip -6mm d\tau}\, V_2(\tau) \, \cV_3(0) \, \cV_4(-\infty)\Big\r\;,\\
    \cA_{1234}^{\mathrm{II}}&= i\,\Big\l\cV_1(+\infty) \, \cV_{23}(0)\, \cV_4(-\infty)
\Big\r\;, \\
    \cA_{1234}^{\mathrm{III}}&= i\,\Big\l\cV_1(+\infty) \, \cV_{23}^{\rm pinch}(0)\, \cV_4(-\infty) \Big\r\;.
\end{align}
As we have discussed in the previous section, the second contribution $\cA_{1234}^{\mathrm{II}}$ comes from the insertion of the quadratic vertex operator
\begin{equation}\label{V23}
\cV_{23}(\tau)=4\,c(\tau)\,\epsilon^{[\mu}_{2}\epsilon_{3}^{\nu]}\alpha_\mu(\tau)\,\bar\alpha_\nu(\tau)\,e^{i(k_2+k_3)\cdot x(\tau)}   \;,
\end{equation}
corresponding to the quartic vertex. The third contribution, instead, comes from ``pinching'' the gluons 2 and 3 into a cubic subtree attached to the worldline. This is done by inserting an unintegrated vertex operator \eqref{Vertex operators}, corresponding to a nonlinear field with momentum $k_{23}^\mu=k^\mu_2+k^\mu_3$ and composite polarization
\begin{equation}
\epsilon_{23}^\mu=\frac{1}{k_{23}^2}\,m_2^\mu(\epsilon_2,\epsilon_3)\;,   
\end{equation}
written in terms of the product \eqref{C-inf prod}.

The necessity for the second and third contributions to the amplitude can be understood from studying its Ward identities. To this end, we take the first gluon to have the pure gauge polarization $\epsilon_1^\mu=ik_1^\mu$, so that $\cV_1=[Q,e^{ik_1\cdot x}]$. Following the procedure described before, the three correlation functions contribute as
\begin{equation}
\begin{split}
\cA_{k_1234}^{\mathrm{I}}&=i\int_{0}^{\infty}\!\!\!\!d\tau\,\Big\l e^{ik_1\cdot z(+\infty)}\,\frac{d}{d\tau}\cV_2(\tau) \, \cV_3(0) \, \cV_4(-\infty)\Big\r=-i\,\Big\l e^{ik_1\cdot z(+\infty)}\,\cV_2(0) \, \cV_3(0) \, \cV_4(-\infty)\Big\r\;,\\
\cA_{k_1234}^{\mathrm{II}}&= -i\,\Big\l e^{ik_1\cdot z(+\infty)} \,\{Q,\cV_{23}(0)\}\, \cV_4(-\infty)
\Big\r\;,\\
\cA_{k_1234}^{\mathrm{III}}&= -i\,\Big\l e^{ik_1\cdot z(+\infty)} \,\{Q,\cV_{23}^{\rm pinch}(0)\}\, \cV_4(-\infty)
\Big\r\;.
\end{split}    
\end{equation}
Compared to the smart worldline, the first correlator does not vanish, since the $\tau$ integral receives a boundary contribution from $\cV_2(\tau)$ colliding with $\cV_3(0)$. Instead of computing the full correlation functions, we will determine the OPE $\cV_2(0)\cV_3(0)$ at coinciding points, and show that it cancels $\{Q,\cV_{23}(0)+\cV_{23}^{\rm pinch}(0)\}$. 

For later convenience, let us rewrite the unintegrated vertex operator in the compact form
\begin{equation}
\cV_i=\epsilon^\mu_{i}\big(S_\mu+i\,c\,\dot x_\mu\big)e^{ik_i\cdot x}+2i\,c\,k^\mu_i\epsilon^\nu_i\,S_{\mu\nu}\,e^{ik_i\cdot x}\;,    
\end{equation}
where the time dependence is left implicit and we have introduced the ghost-number one vector $S^\mu$ and the Lorentz spin generator $S^{\mu\nu}$, defined by
\begin{equation}
S^\mu=\bar\cC\alpha^\mu+\cC\bar\alpha^\mu\;,\quad S^{\mu\nu}=\alpha^\mu\bar\alpha^\nu-\alpha^\nu\bar\alpha^\mu \;.   
\end{equation}
Notice that, for $k_i^2=0$ and $k_i\cdot\epsilon_i=0$, the vertex operator is normal ordered.
In computing the OPE, we will leave equal-time products of $\dot x^\mu$ and functions of $x^\mu$ as they are, without taking contractions. This corresponds to the Weyl ordering of the related operators. For instance, the product $e^{ik_i\cdot x(\tau)}e^{ik_j\cdot x(\tau)}=e^{i(k_i+k_j)\cdot x(\tau)}$ is Weyl ordered, not normal ordered, unless the propagator $G(\tau,\tau)$ vanishes. For equal-time products of $\alpha^\mu$ and $\bar\alpha^\nu$ we will use instead the OPE at coincident times
\begin{equation}
\bar\alpha^\mu(\tau)\,\alpha^\nu(\tau)=\frac12\,\eta^{\mu\nu}+:\bar\alpha^\mu(\tau)\,\alpha^\nu(\tau):    \;,
\end{equation}
to take advantage of the fact that normal ordered expressions with more than one barred field vanish in all correlation functions due to charge conservation, c.f. \eqref{Normal order vanishing}. Some useful examples are
\begin{equation}
\begin{split}
S^\mu S^\nu&=-:\cC\bar\cC(\alpha^\mu\bar\alpha^\nu-\alpha^\nu\bar\alpha^\mu):\;=0 \;,\\
S^{\mu\nu}S^\rho&=\tilde S^{[\mu}\eta^{\nu]\rho}+:S^{\mu\nu} S^\rho:\;= \tilde S^{[\mu}\eta^{\nu]\rho}\;,
\end{split}    
\end{equation}
where in the final results we have discarded all terms vanishing inside correlators, and we have defined a second ghost-number one vector
\begin{equation}
\tilde S^\mu= \bar\cC\alpha^\mu-\cC\bar\alpha^\mu\;.   
\end{equation}

Armed with these relations, we can determine the equal-time OPE of the two vertex operators $\cV_2(0)$ and $\cV_3(0)$. Omitting the time dependence, we have
\begin{equation}
\begin{split}
\cV_2\cV_3&=\Big(\epsilon^\mu_{2}\big(S_\mu+i\,c\,\dot x_\mu\big)+2i\,c\,k^\mu_2\epsilon^\nu_2\,S_{\mu\nu}\Big)\Big(\epsilon^\rho_{3}\big(S_\rho+i\,c\,\dot x_\rho\big)+2i\,c\,k^\rho_3\epsilon^\sigma_3\,S_{\rho\sigma}\Big)e^{ik_{23}\cdot x} \\
&=2i\,c\Big(\epsilon^{[\mu}_{2}\epsilon^{\nu]}_{3}\dot x_\mu S_\nu+k_2^{[\mu}\epsilon_2^{\nu]}\tilde S_\mu\epsilon_{3\nu}-k_3^{[\mu}\epsilon_3^{\nu]}\tilde S_\mu\epsilon_{2\nu} \Big)e^{ik_{23}\cdot x}\\
&=c\,\tilde S_\mu\,m_2^\mu(\epsilon_2,\epsilon_3)\,e^{ik_{23}\cdot x}+2i\,c\,\epsilon^{[\mu}_{2}\epsilon^{\nu]}_{3}\Big(\dot x_\mu S_\nu+\tilde S_\mu k_{23\nu}\Big)e^{ik_{23}\cdot x}\;,
\end{split}    
\end{equation}
The quadratic vertex operator \eqref{V23} can be written as
\begin{equation}
\cV_{23}=2\,c\,\epsilon^\mu_2\epsilon^\nu_3\,S_{\mu\nu}\,e^{ik_{23}\cdot x}\;,
\end{equation}
and its equal-time commutator with $Q$ can be computed as $\{Q,\cV_{23}\}=is(\cV_{23})$, where the BRST differential $s$ on the worldline variables is given by \eqref{BRST transf lagrangian}. Its action on functions is defined by the Leibniz rule, with the resulting expressions corresponding to Weyl ordering. Discarding again all terms with two barred fields in normal ordering, we obtain
\begin{equation}
is(\cV_{23})=-2i\,c\,\epsilon^{[\mu}_{2}\epsilon^{\nu]}_{3}\Big(\dot x_\mu S_\nu+\tilde S_\mu k_{23\nu}\Big)e^{ik_{23}\cdot x}  \;,  
\end{equation}
yielding the OPE result
\begin{equation}\label{OPE intermediate}
\cV_2\cV_3+\{Q,\cV_{23}\}=c\,\tilde S_\mu\,m_2^\mu(\epsilon_2,\epsilon_3)\,e^{ik_{23}\cdot x} \;,   
\end{equation}
where all operators above are evaluated at equal times. 

At this stage, the missing ingredient to establish the Ward identity is to prove that the right-hand side of \eqref{OPE intermediate} is $Q$-exact and, in particular, given by
\begin{equation}\label{final Ward}
c\,\tilde S_\mu\,m_2^\mu(\epsilon_2,\epsilon_3)\,e^{ik_{23}\cdot x}=-\{Q,\cV_{23}^{\rm pinch}\}\;.   
\end{equation}
To this end, we consider the vertex operator for an off-shell field
\begin{equation}\label{Voffshell}
\cV_A=A^\mu(k)\big(S_\mu+i\,c\,\dot x_\mu\big)e^{ik\cdot x}+2i\,c\,k^\mu A^\nu(k)\,S_{\mu\nu}\,e^{ik\cdot x}\;,    
\end{equation}
where in general $k^2\neq0$ and $k\cdot A(k)\neq0$. Its $Q$-commutator is readily computed, yielding
\begin{equation}\label{MaxwellQ}
\{Q,\cV_A\}=c\,\tilde S^\mu\big(-k^2A_\mu+k_\mu k\cdot A\big)\,e^{ik\cdot x} \;.   
\end{equation}
For the case at hand, $\cV_{23}^{\rm pinch}\equiv\cV_{A_{23}}$ is of the form \eqref{Voffshell} for a composite field $A_{23}^\mu$, where the momentum $k_{23}^\mu=k_2^\mu+k_3^\mu$ is off-shell and the polarization is given by
\begin{equation}
\epsilon_{23}^\mu=\frac{1}{k_{23}^2}\,m_2^\mu(\epsilon_2,\epsilon_3)\equiv\frac{i}{k_{23}^2}\,\big(\epsilon_2\cdot k_3\,\epsilon_3^\mu-\epsilon_3\cdot k_2\,\epsilon_2^\mu+(k_2-k_3)^\mu\,\epsilon_2\cdot\epsilon_3\big)\;.
\end{equation}
One can easily check that $k_{23}\cdot m_2(\epsilon_2,\epsilon_3)=0$ (this is essentially the Ward identity for the three-point case), so that \eqref{MaxwellQ} yields
\begin{equation}
\begin{split}
\{Q,\cV_{23}^{\rm pinch}\}&=c\,\tilde S_\mu\big(-k_{23}^2\epsilon_{23}^\mu+k_{23}^\mu k_{23}\cdot\epsilon_{23}\big)\,e^{ik_{23}\cdot x}=c\,\tilde S_\mu\big(-k_{23}^2\epsilon_{23}^\mu\big)\,e^{ik_{23}\cdot x}\\
&=-c\,\tilde S_\mu\,m_2^\mu(\epsilon_2,\epsilon_3)\,e^{ik_{23}\cdot x}\;,    
\end{split}
\end{equation}
thus proving \eqref{final Ward} and hence the Ward identity. 

Let us conclude by emphasizing that the above discussion demonstrates how the form of the composite field \eqref{A23} can also be inferred (albeit in a somewhat roundabout way) by demanding the Ward identities. Moreover, as expected, the Ward identity remains unchanged whether one uses $A_{23}^\mu$ or its gauge-transformed version $\tilde A_{23}^\mu$ \eqref{BGtilde}, since the two fields differ by a gauge transformation:
\begin{equation}
\{Q,\cV_{\tilde A_{23}}\}=\{Q,\cV_{A_{23}}\}+\{Q,[Q,\lambda_{23}]\}=\{Q,\cV_{A_{23}}\}\;,    
\end{equation}
where $\lambda_{23}(x)$ is given by \eqref{composite lambda} and we have used $\{Q,Q\}=0$ together with the graded Jacobi identity of (anti)commutators.

\section{Conclusions and Outlook} \label{sec5}
In this paper, we have quantized the bosonic spinning particle on the open worldline, with the aim of providing a first-quantized description of gluon scattering amplitudes. To this end, we considered an open worldline of infinite proper length \cite{Daikouji:1995dz,
Laenen:2008gt, Bonocore:2020xuj,Mogull:2020sak}, with boundary conditions corresponding to zero-momentum asymptotic vacua. In this way, all external gluons (including the two at the asymptotic endpoints of the line) are described by insertions of suitable vertex operators, thus mimicking the worldsheet techniques of string perturbation theory. In particular, the infinite length of the line allowed us to bypass the LSZ procedure \cite{Mogull:2020sak,Bonezzi:2025iza}, in that the correlation functions already provide the on-shell and amputated amplitudes. Contrary to string theory, however, linear vertex operators are not sufficient to compute arbitrary amplitudes. In general, one needs nonlinear vertex operators, describing the insertion of entire subtrees on the main worldline. While this is typically achieved by using the so-called Bern-Kosower pinching rules \cite{Bern:1990cu,Bern:1991an,Bern:1992ad,Schubert:2001he}, here we have argued that these nonlinear vertex operators can be obtained as well from worldline path integrals on a semi-infinite open geometry \cite{Bonezzi:2025iza}. Finally, to test our results, we have shown how the Ward identities of the scattering amplitudes descend from the BRST invariance of the worldline correlators.

The framework presented in this work should be extended to arbitrary multiplicities and, possibly, to loop amplitudes. In particular, one would like to be able to prove gauge invariance of arbitrary amplitudes by extending the BRST analysis introduced here to general correlation functions. Besides the point of principle, having a first-quantized description of Yang-Mills amplitudes could help in shedding light on the so-called duality between color and kinematics in gauge theories \cite{Bern:2008qj,Bern:2019prr}, which underpins the double copy construction of gravity amplitudes \cite{Bern:2010ue,Bern:2022wqg,Adamo:2022dcm}. More specifically, it would prove beneficial to connect the off-shell algebraic approach of \cite{Bonezzi:2022bse,Bonezzi:2024emt,Bonezzi:2024fhd} to the worldline techniques employed in \cite{Ahmadiniaz:2021fey,Ahmadiniaz:2021ayd}, where an algorithm combining worldline integration by parts and Bern-Kosower pinching rules was shown to produce gauge theory numerators obeying color-kinematics duality.
Since the bosonic spinning particle can also accommodate gravitons in its spectrum, it would be interesting to extend the present framework to the spin-two sector of the theory. There, one expects to need multi-graviton vertices of arbitrary order, but the linear ones should essentially be the product of two gluon vertex operators, thus showing a worldline incarnation of the double copy, along the lines of \cite{Shi:2021qsb,Bastianelli:2021rbt}.

\section*{Acknowledgments}

We would like to thank Christoph Chiaffrino, Olaf Hohm and Maria Kallimani for useful discussions. 
We are grateful to the organizers of the workshop ``New Trends in First Quantization: Field Theory, Gravity and Quantum Computing" (Physikzentrum Bad Honnef) for providing an environment that inspired fruitful discussions from which this research benefited. The work of R.B. is funded by the Deutsche Forschungsgemeinschaft (DFG, German
Research Foundation) – Projektnummer 524744955, “Worldline approach to the double
copy”.

\appendix

\section{Path integral in phase space} \label{Appendix:A}
To study the propagator directly in momentum space, we wish to consider the path integral representation of the matrix element 
 \begin{equation}
\la p' | e^{-i HT} |p\ra = \int Dp Dx \  e^{iS}\;,
\label{app-pi} 
\end{equation}
where one could take $H= p^2$ as needed in the main text. We study the simpler  case 
$H=0$, which already captures all the essential aspects of our discussion. 
For simplicity, we consider just one target space dimension. 
Then, the left-hand side of \eqref{app-pi} reduces to 
\be
\la p' |p\ra = 2\pi \delta (p-p')\;,
\label{app-2}
\ee
which should be reproduced by the path integral on the right-hand side. To take into account the 
boundary conditions, that should be imposed on $p(\tau)$ only, the action takes the form
\be
S =  \int_0^T d\tau\,(- \dot p x ) \;.
\ee
Of course, one could perform the path integral by first integrating over $p$ and then over $x$, which works well 
for nontrivial quadratic Hamiltonians, like the free one, $H= p^2$. This is well-known.
Here we wish to path integrate in $x$ first, leaving the path integral on $p$ as the last one. 
To do this, we notice that the paths $x(\tau)$ should not contain any boundary conditions, so that we can 
expand them by extracting the constant path $x_0$. That is, we parametrize
\be
x(\tau)= x_0 +\tilde x(\tau) 
\ee 
with $\int_0^T d\tau\, \tilde x(\tau) =0$. 
Then, we can write
\be
\int Dp\, D\tilde x\, dx_0 \  e^{-i\int_0^T \! d\tau\,  \dot p x} = \int Dp\,  D\tilde x\, 
\underbrace{dx_0\,
\  e^{-i x_0 \int_0^T \! d\tau\,  \dot p}}_{\sim \delta(p-p')}
 e^{ -i \int_0^T \! d\tau\, \dot p  \tilde x  }\;,  
 \ee
where $p$ and $p'$ denote the boundary conditions that must be imposed on $p(\tau)$.
Of course, $\int_0^T \! d\tau\,  \dot p = p'-p$.
Then, path integrating over $\tilde x(\tau)$ produces a delta functional $\delta(\dot p) $
 leading to
\be
\int Dp\, D x \  e^{-i\int_0^T \! d\tau\, \dot p x} 
\ \sim \
\delta(p-p')
\int Dp\, \delta(\dot p) 
\ \sim \ 
\text{Det}(\partial_\tau) 
\delta(p-p') 
\ \sim \ 
2\pi \delta(p-p')\;,
 \ee
where the combined unfixed constants have been chosen to match the expected result in \eqref{app-2}.
This way of computing the path integral allows us to bypass configuration space and obtain 
results directly in momentum space. Presumably, perturbation theory can be introduced by building further 
on the above expression. 

Let us also briefly outline the derivation of the other ``phase-space" path integral involved in our derivation, the one for the bosonic oscillators, 
\begin{align} \label{A.7}
Z(\alpha,\bar\alpha,\theta)=\int_{\bar\alpha(0)=\bar\alpha}^{\alpha(1)=\alpha} \!\!\!D\bar\alpha D\alpha\ e^{\int_0^1d\tau\, \bar\alpha^\mu(\partial_\tau-i\theta)\alpha_\mu(\tau)+\bar\alpha^\mu\alpha_\mu(0)}= e^{e^{-i\theta}\alpha\cdot\bar\alpha}\ ,
\end{align}
which was, for example, discussed and used in Ref. \cite{Ahmadiniaz:2015xoa}. It corresponds to the usual (Euclidean) path integral for the harmonic oscillator, though with an imaginary frequency $\omega= - i\theta$.

For $\theta=0$, the latter reduces to the path integral representation of the scalar product of the coherent states,
\begin{align}
    \langle \alpha|\bar\alpha\rangle = e^{\alpha\cdot \bar\alpha} = \int_{\bar\alpha(0)=\bar\alpha}^{\alpha(1)=\alpha} \!\!\!D\bar\alpha D\alpha\ e^{\int_0^1d\tau\, \bar\alpha^\mu\dot\alpha_\mu(\tau)+\bar\alpha^\mu\alpha_\mu(0)}\;,
\end{align}
which can be used to fix the path integral normalization as follows. Let us split the fields into  background parts -- which satisfy the equations of motion $\dot\alpha^\mu =\dot{\bar\alpha}^\mu=0$ and the boundary conditions -- and quantum fluctuations as
\begin{align}
    &\alpha^\mu(\tau)=\alpha^\mu +\kappa^\mu(\tau)\,,\quad \kappa^\mu(1)=0\;,\\
&\bar\alpha^\mu(\tau)=\bar\alpha^\mu +\bar\kappa^\mu(\tau)\,,\quad \bar\kappa^\mu(0)=0\ .
\end{align}
Thus, we get
\begin{align}
    e^{\alpha\cdot \bar\alpha} =e^{\alpha\cdot \bar\alpha} \int_{\bar\kappa(0)=0}^{\kappa(1)=0} \!\!\!D\bar\kappa D\kappa\ e^{\int_0^1d\tau\, \bar\kappa^\mu\dot\kappa_\mu(\tau)}\ ,
\end{align}
i.e.,
\begin{align}
\int_{\bar\kappa(0)=0}^{\kappa(1)=0} \!\!\!D\bar\kappa D\kappa\ e^{\int_0^1d\tau\, \bar\kappa^\mu\dot\kappa_\mu(\tau)}=1\ .
\label{eq:PI-normalization}
\end{align}
We can now apply the background splitting to the $\theta$-dependent path integral. We have,
\begin{align}
&\alpha^\mu(\tau)=e^{i\theta\tau} \big(e^{-i\theta} \alpha^\mu +\kappa^\mu(\tau)\big)\,,\quad \kappa^\mu(1)=0\;,\\
&\bar\alpha^\mu(\tau)=e^{-i\theta\tau} \big(\bar\alpha^\mu +\bar\kappa^\mu(\tau)\big)\,,\quad \bar\kappa^\mu(0)=0\ ,
\end{align}
and, in turn,
\begin{align}
Z(\alpha,\bar\alpha,\theta)=e^{e^{-i\theta}\alpha\cdot\bar\alpha}\int_{\bar\kappa(0)=0}^{\kappa(1)=0} \!\!\!D\bar\kappa D\kappa\ e^{\int_0^1d\tau\, \bar\kappa^\mu\dot\kappa_\mu(\tau)} = e^{e^{-i\theta}\alpha\cdot\bar\alpha}\ .
\label{A.15}
\end{align}
One may consult Ref. \cite{Ahmadiniaz:2015xoa} for further details.


\bibliography{gluons.bib}
\bibliographystyle{utphys}


\end{document}